\documentclass[a4paper,fleqn,usenatbib]{mnras}
\usepackage{newtxtext,newtxmath}
\usepackage[T1]{fontenc}
\usepackage{ae,aecompl}

\usepackage[latin1]{inputenc}
\usepackage[]{graphicx}	% Including figure files
\usepackage[hyphenbreaks]{breakurl}
\usepackage{amsmath}	% Advanced maths coommands
\usepackage{amssymb}	% Extra maths symbols

\usepackage[lofdepth,lotdepth]{subfig}
\usepackage{lscape}
\usepackage{placeins}
\usepackage{xspace}

\defcitealias{paper1}{paper I\xspace}
\newcommand{\paperI}{\citetalias{paper1}\xspace}
\newcommand{\paperIp}{\citepalias{paper1}\xspace}

\DeclareRobustCommand{\Chi}{{\mathpalette\irchi\relax}}
\newcommand{\irchi}[2]{\raisebox{\depth}{$#1\chi$}}

\newcommand{\refeq}[1]{Eq. (\ref{#1})}
\newcommand{\refeql}[1]{(see Eq. \ref{#1})}

\newcommand{\refsec}[1]{Sec. \ref{#1}}
\newcommand{\refsecl}[1]{(see Sec. \ref{#1})}
\newcommand{\refim}[1]{Fig. \ref{#1}}
\newcommand{\refapp}[1]{Appendix \ref{#1}}
\DeclareMathOperator\arctanh{arctanh}

\newcommand{\avg}[1]{\left\langle #1 \right\rangle}
\newcommand{\abs}[1]{\left| #1 \right|}
\newcommand{\order}[1]{\mathcal{O}\left(#1\right)}
\newcommand{\Mpc}{$h^{-1}$Mpc\xspace}
\newcommand{\Msun}{$h^{-1}M_\odot$\xspace}

\newcommand{\ie}{i.e.\xspace}

\newcommand{\J}{\left\langle J_1 \right\rangle}
\newcommand{\kkJ}{\left\langle k^2J_1 \right\rangle}
\newcommand{\JJ}{\left\langle J_1^2 \right\rangle}
\newcommand{\W}{\left\langle W_1 \right\rangle}
\newcommand{\kkW}{\left\langle k^2W_1 \right\rangle}
\newcommand{\WW}{\left\langle W_1^2 \right\rangle}
\newcommand{\WJ}{\left\langle W_1J_1 \right\rangle}
\newcommand{\kk}{\left\langle k^2 \right\rangle}
\newcommand{\kkkk}{\left\langle k^4 \right\rangle}
\newcommand{\Jnm}[2]{\mathcal{J}_{#1,#2}}

\newcommand{\LCDM}{$\Lambda$CDM\xspace}

\newcommand{\varsc}{\mathcal{R}}
\newcommand{\vart}{\tau}
\newcommand{\bx}{\boldsymbol{x}}

\makeatletter
\let\@fnsymbol\@arabic
\makeatother

\title[Statistical properties of CoSpheres]{Probability distribution and statistical properties of spherically compensated cosmic regions in $\Lambda$CDM cosmology}

\author[Jean-Michel Alimi $\&$ Paul de Fromont ]{Jean-Michel Alimi\textsuperscript{\thanks{email:jean-michel.alimi@obspm.fr}} Paul de Fromont\textsuperscript{\thanks{email:paul.de-fromont@obspm.fr}}\\
	LUTH, Observatoire de Paris, PSL Research University, CNRS, Université Paris Diderot, Sorbonne Paris Cité\\
	5 place Jules Janssen, 92195 Meudon}

\date{Accepted XXX. Received YYY; in original form ZZZ}

\pubyear{2017}

\begin{document}
\label{firstpage}
\pagerange{\pageref{firstpage}--\pageref{lastpage}}
\maketitle

\begin{abstract}
The statistical properties of cosmic structures are well known to be strong probes for cosmology. In particular, several studies tried to use the cosmic void counting number to obtain tight constrains on Dark Energy. In this paper we address this question by using the CoSphere model as introduced in \citet{paper1}. We derive their exact statistics in both primordial and non linearly evolved Universe for the standard \LCDM model. We first compute the full joint Gaussian probability distribution for the various parameters describing these profiles in the Gaussian Random Field. We recover the results of \citet{BBKS} only in the limit where the compensation radius becomes very large, i.e. when the central extremum decouples from its cosmic environment. We derive the probability distribution of the compensation size in this primordial field. We show that this distribution is redshift independent and can be used to model cosmic void size distribution. Interestingly, it can be used for central maximum such as DM haloes. We compute analytically the statistical distribution of the compensation density in both primordial and evolved Universe. We also derive the statistical distribution of the peak parameters already introduced by \citet{BBKS} and discuss their correlation with the cosmic environment. We thus show that small central extrema with low density are associated with narrow compensation regions with a small $R_1$ and a deep compensation density $\delta_1$ while higher central extrema are located in larger but smoother over/under massive regions. 
\end{abstract}


%%%%%%%%%%%%%%%%
%%%% INTRODUCTION %%%
%%%%%%%%%%%%%%%%%
\section*{Introduction}
Statistical properties of high density regions (as dark matter (DM) haloes) or under dense regions (as cosmic voids) have been extensively used to address the main questions of modern cosmology such as the origin of dark energy (DE) or the nature of gravity. Numerous successes have been obtained from the mass function of DM haloes through the Press Schechter formalism \citep{Press1974} or its powerful extensions like Excursion Set Theory \citep{Bond1991}. Predictions using these formalism are generally in very good agreement with numerical simulation results \citep{Sheth1999, Jenkins2001, Tramonte2017} but these formalisms do not probe the large scale environment of DM haloes. Moreover a full understanding of such cosmological probes needs a full or at least a better understanding of the non linear evolution of gravitational collapse. 

Concerning under dense regions as cosmic voids, it is even more challenging to describe precisely the statistics of such regions \citep{Sheth2004}, mainly because we do not have an objective definition and a physically motivated dynamical model for voids. Both dynamical and statistical properties of cosmic voids depend on their algorithmic definition \citep{Platen2007, Neyrinck2008, Cautun2016}, a full comparative analysis of algorithms for detecting voids in numerical simulations is for example necessary. 

In \citet{paper1}, labelled thereafter \paperI, we introduced the spherically compensated cosmic regions, named thereafter CoSpheres. Such regions describe the large scale cosmic environment around local extremum in the density field. CoSphere can be splitted in two distinct radial regions. An over (resp under) massive spherical core around the central maximum (resp minimum) and an exterior under (resp over) massive surrounding belt. By over massive we mean that the total mass $m(r)$ is higher than the homogeneous mass $4\pi/3\bar{\rho}_m r^3$. In the Newtonian limit, over massive regions collapse (\ie $\ddot{r}<0$) while under massive region expand toward larger radii. For each central extremum, the radius separating these two distinct regions is called the compensation radius $R_1$. By definition, it satisfies $m(R_1)=4\pi/3\bar{\rho}_m R_1^3$. The origin of CoSpheres within the primordial Gaussian Random Field (GRF) has been precisely described using the constrained GRF formalism with an appropriate compensation constraint \paperIp. In this primordial Gaussian field, the expected spherically average profiles can be fully parametrized by four independent scalars. Beside the compensation radius $R_1$, they are described by three \textit{shape} parameters: $\nu$, $x$ and $\nu_1$. The first parameters $x$ and $\nu$, already introduced by \cite{BBKS}, qualify the central extrema while $\nu_1$ defines the compensation density contrast $\delta_1=\nu_1\sigma_0$ as $\delta(R_1)=\delta_1$.

The non linear dynamical evolution of CoSpheres is described with high precision through the spherical collapse model. These cosmic regions can be detected in numerical simulations, in \paperI we showed that they can be fully reconstructed from high redshift (within the Gaussian random field) until $z=0$ in $\Lambda$CDM cosmology. Consequently, these regions can be used as powerful probes for cosmology and gravity itself as it will be investigated in \citet{paper3, paper4}. 
 
While \paperI focused on the construction of these cosmic regions and the derivation of their average density and mass profiles at any redshift, this paper is fully dedicated to the study of their statistical properties. We thus derive the full joint Gaussian probability distribution for the profile parameters $R_1$, $\nu$, $x$ and $\nu_1$ in GRF. This distribution measures the probability to obtain a CoSphere with the corresponding parameters in the primordial Gaussian Universe.

We then deduce the one-dimensional probability distribution $dP(R_1)$ marginalized over the shape parameters $\nu$, $x$ and $\nu_1$. This distribution is proportional to the count number of compensation radii. It gives the probability to find a $R_1$ around any extremum. Despite being derived in the primordial Gaussian field, since compensation radii evolve comovingly \paperIp, this distribution is expected to be redshift-independent. Using numerical simulations, we show that it is indeed well conserved during evolution. Interestingly, this size distribution provides a well defined analytical prediction for cosmic voids sizes once considered as compensated regions around minimum whose size is defined as $R_1$. 

From the full joint Gaussian probability distribution, we also compute the marginalized conditional distribution of the three shape parameters at a given compensation radius $R_1$. We then derive their constrained moments $\avg{\alpha^n|R_1}$ with $\alpha=\{\nu,x,\nu_1\}$. For $n=1$, the mean values $\avg{\nu|R_1}$, $\avg{x|R_1}$ and $\avg{\nu_1|R_1}$ can be used to define the mean average profile at fixed compensation radius. These profiles are expected to reproduce the full matter field of CoSpheres once averaged over all possible stochastic realization, \ie all possible value for each shape parameters $\nu$, $x$ and $\nu_1$ given $R_1$. We then study the shape of the mean average profiles according to $R_1$ and show that the central extrema progressively tends to the universal BBKS peak profile \citep{BBKS} for large $R_1$. For small $R_1$ however, the central extremum is strongly correlated to its cosmic environment through $\nu_1$ and $R_1$. 

Using the spherical collapse model, we derive the exact non linear evolution of the compensation density distribution $dP(\delta_1, R_1)$ for any redshift. We compute analytically the evolved moments $\avg{\delta_1^n|R_1}$ for both cosmic voids (central minimum) and central over densities. We compare our results with numerical simulation and show that the agreement is very  good, even in the non linear regime.

This paper is organized as follow : in the first section we define precisely CoSpheres and their compensation radius $R_1$. We also discuss how such cosmic regions are detected in numerical simulation. In \refsec{initial conditions} we derive the statistical properties of these regions in the primordial Gaussian field, the radii distribution $dP(R_1)$ together with the statistical study of the shape parameters. We discuss the properties of the mean averaged density profile at fixed compensation radius $R_1$. In the last section, \refsec{dynamics}, we study the dynamical properties of these distribution by using the Lagrangian Spherical Collapse and compare the results to numerical simulations at $z=0$ in \LCDM cosmology.

%%%%%%%%%%%%%%%%
%%%%%   SECTION 1 %%%%
%%%%%%%%%%%%%%%%%
\section{CoSpheres in the sky}
\label{section1}
We study the statistical properties of CoSpheres. These cosmic structures are defined around extrema (minima or maxima) in the density field at any redshift \paperIp. Around each extremum we define the concentric mass $m(r)$ as the mass enclosed in the sphere of radius $r$, from which we deduce the spherical mass contrast $\Delta(r)$ as
\begin{equation}
\label{Delta}
\Delta(r):=\frac{m(r)}{4\pi/3\bar{\rho}_mr^3} - 1
\end{equation}
This profile is linked to the density contrast $\delta(r)=\rho_m(r)/\bar{\rho}_m - 1$ through
\begin{equation}
\label{delta_vs_Delta}
\Delta'(r)=\frac{3}{r}\left[\delta(r)-\Delta(r)\right]\Leftrightarrow \Delta(r)=\frac{3}{r^3}\int_0^ru^2\delta(u)du
\end{equation}
where $\Delta'(r) = \partial \Delta (r)/ \partial r$. As discussed in \paperI, each extremum must be compensated on a finite scale. For each spherical profile, it exists a unique scale $R_1$ called \textit{compensation radius} satisfying. 
\begin{equation}
\label{R1}
\Delta(R_1) = 0
\end{equation}
$R_1$ is defined as the smallest radius satisfying \refeq{R1}. This scale measures the size of the over (resp. under) massive region\footnote{not to be confused with over/under-dense regions} surrounding each maximum (resp. minimum). Since $\ddot{r}\propto - \Delta(r)$ in Newtonian regime, the mass contrast $\Delta(r)$ drives the local gravitational collapse. The compensation radius separates the collapsing and the expanding regions.

\begin{figure*}
	\captionsetup[subfigure]{width=0.48\linewidth}
	\begin{center}
		\subfloat[Average profiles around haloes with a mass $M_h\sim~3.0\times 10^{13}$ \Mpc at $z=0$.]{
			\includegraphics[width=0.5\textwidth]{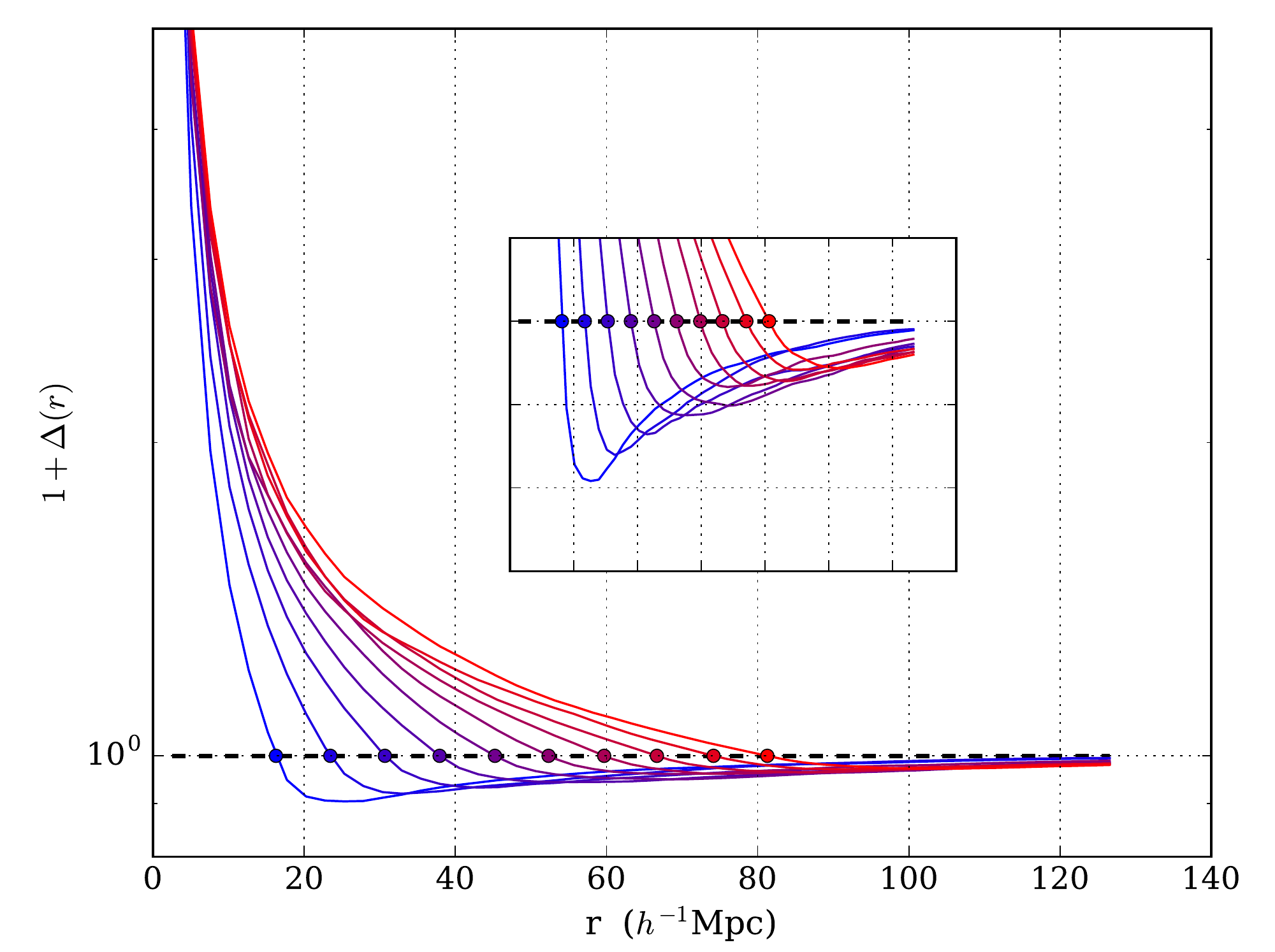}
			\label{fig:profile_D_today}}
		\subfloat[Same as in left panel for central minimum, \ie cosmic voids.]{
			\includegraphics[width=0.5\textwidth]{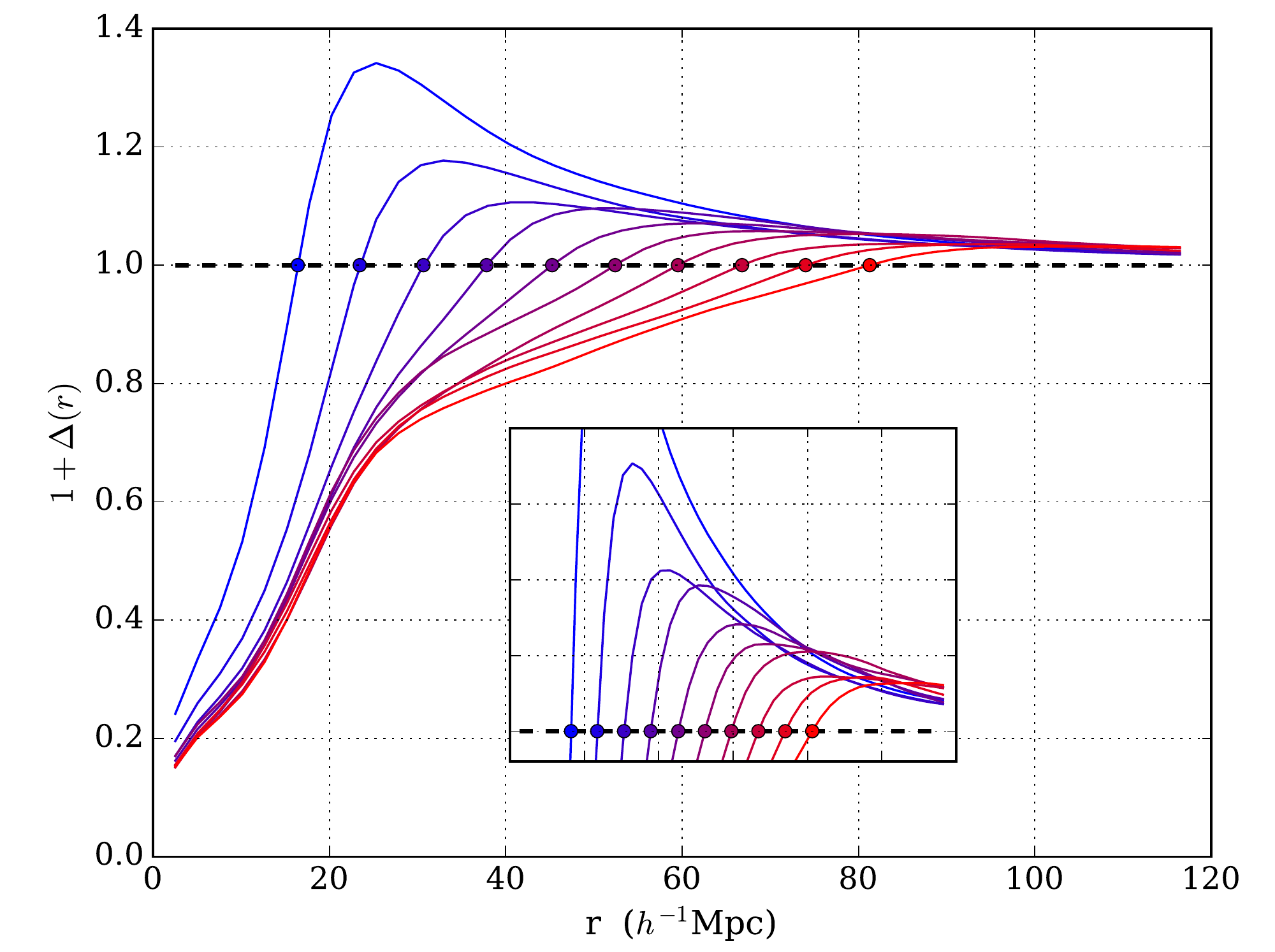}
			\label{fig:profile_D_void_today}}
		\caption{Radial average mass contrast at $z = 0$ in the reference simulation. Each curve corresponds to a given compensation radius $R_1$ from $15$ to $80$ \Mpc. Whereas each single individual profile is far from a smooth curve, stacked profiles display a global shape with well defined properties.} 
		\label{fig:averaged_profile}
	\end{center}
\end{figure*}

These regions can be detected in numerical simulations. We use in this work, the numerical simulations from the ``Dark Energy Universe Simulation'' (DEUS) project, publicly available through the ``Dark Energy Universe Virtual Observatory ''  (DEUVO) Database \footnote{http://www.deus-consortium.org/deus-data/}. These simulations consist of N-body simulations of Dark Matter (DM) for realistic dark energy models. For more details we refer the interested reader to dedicated sections in \citet{Alimi2010, Rasera2010, Courtin2010, Alimi2012, Reverdy2015}. We focus in this paper on the flat $\Lambda$CDM model with parameters calibrated against measurements of WMAP 5-year data \citep{Komatsu2009} and luminosity distances to Supernova Type Ia from the UNION dataset \citep{Kowalski2008}. 

The reduced Hubble constant is set to $h=0.72$ and the cosmological parameters are $\Omega_{DE}=0.74$, $\Omega_b=0.044$, $n_s=0.963$ and $\sigma_8=0.79$. All along this work, The \textbf{reference simulation} is chosen with $L_{box}=2592$ $h^{-1}Mpc$ and $n_{part}=2048^3$. It provides both a large volume and a good mass resolution. Here the mass of one particle is $m_p\sim 1.5\times 10^{11}$ \Msun.

The construction procedure of numerical CoSpheres consists first in finding the position of local extremum. In the case of a central over-density we identify maxima with the center of mass of DM haloes. Halos are founded by a Friend-of-Friend algorithm with a linking length $b=0.2$. We considered in the reference simulation $200 000$ haloes with a mass $M_h\sim 3\times 10^{13}$ \Msun. Selecting haloes with a mass $M_h$ is equivalent to impose a threshold on the height of their progenitor, \ie it selects local extrema with $\nu\geq \nu_0=\delta_c/\sigma_0(M_h)$ where $\delta_c\simeq 1.686$ for $\Lambda$CDM cosmology and $\sigma_0(M_h)$ is the fluctuation level. 

For central under-densities we smooth the density field with a Gaussian kernel on a few number of cells. Minima are founded by comparing the local density of each cell to its neighbours. The center of the cell is then identified with the position of the local minimum. The backward procedure is simplistic and assumes that the comoving position of each void is conserved during cosmic evolution. At any redshift, each void's position is assumed to be the same than the one detected at $z=0$.

From each extrema, we compute the concentric mass $m(r)$ from DM particles
\begin{equation}
m(r)=\sum_i m_p\Theta\left[r-\abs{\boldsymbol{x}_i-\boldsymbol{x}_0}\right]
\end{equation}
where $m_p$ is the mass of one particle, $\boldsymbol{x}_i$ the position of the $i^\text{th}$ particle and $\boldsymbol{x}_0$ the position of the central extremum. $\Theta(x)$ is the standard Heaviside distribution such as $\Theta(x)=1$ if $x>0$ and $0$ elsewhere. 

The second step consists into building average profiles by stacking together individual profiles with the same compensation radius. For each $R_1$, we take at least $1000$ profiles for both haloes and voids in order to insure a fair statistics. In \refim{fig:averaged_profile} we show the resulting average profiles for both central over and under densities and several compensation radii at $z=0$ in the reference simulation. As claimed before, the radial structure of these regions is symmetric; a central over (resp. under) massive core until $r=R_1$ surrounded by a large under (resp. over) massive compensation belt for $r\geq R_1$. 

Numerical simulations can be used to follow the gravitational evolution of CoSpheres. By definition, these regions are detected at $z=0$. For a central maximum, \ie build from DM halo, we identify the position of its progenitor at higher redshift to the center of mass of its particles at $z=0$. For each halo detected today, this procedure provides an estimated position of its progenitor at other redshift. These positions are used to define CoSpheres for any $z\neq 0$.

%%%%%%%%%%%%%%%
%%%%  SECTION 2 %%%%
%%%%%%%%%%%%%%%
\section{Statistic of CoSpheres in Gaussian Random Fields}
\label{initial conditions}
In this section we study the statistical properties of CoSpheres in the framework of Gaussian Random Field (GRF) with appropriate constraints \paperIp. 

\subsection{Gaussian Random Fields, the basics}
\label{GRF}

Let us first recall the basic elements necessary for the derivation of average quantities in GRF. We consider here an homogeneous, isotropic random field whose statistical properties are fully determined by its power-spectrum (or spectral density) $P(k)$. It can be written as the Fourier Transform of the auto-correlation of the field $\xi(r)=\xi(|\boldsymbol{x}_1-\boldsymbol{x}_2|)=\avg{ \delta(\boldsymbol{x}_1)\delta(\boldsymbol{x}_2)}$ :
\begin{equation}
\xi(r)=\frac{1}{2\pi^2}\int_0^{+\infty} k^2 P(k) \frac{\sin(kr)}{kr}dk
\end{equation}
The Gaussianity of the field $\delta(\boldsymbol{x})$ leads to the joint probability 
\begin{equation}
d\mathcal{P}_N=P\left[\delta(\boldsymbol{x}_1), ..., \delta(\boldsymbol{x}_N)\right]d\delta(\boldsymbol{x}_1)...d\delta(\boldsymbol{x}_N)
\end{equation}
that the field has values in the range $[\delta(\boldsymbol{x}_i), \delta(\boldsymbol{x}_i) + d\delta(\boldsymbol{x}_i)]$ for each position $\boldsymbol{x}_i$. In this GRF model it is
\begin{equation}
d\mathcal{P}_N=\frac{1}{\sqrt{(2\pi)^N\det \boldsymbol{M}}}\exp\left[-\frac{1}{2}\boldsymbol\delta^t.\boldsymbol{M}^{-1}.\boldsymbol\delta\right]\prod_{i=1}^Nd\delta_i
\end{equation}
$\boldsymbol{\delta}$ is the $N$ dimensional vector $\delta_i=\delta(\boldsymbol{x}_i)$ and $\boldsymbol{M}$ is the $N\times N$ covariance matrix, here fully determined by the field auto-correlation
\begin{equation}
M_{ij}:=\avg{\delta_i\delta_j}=\xi(|\boldsymbol{x}_i-\boldsymbol{x}_j|)
\end{equation}
where the average operator $\avg{...}$ denotes thereafter an ensemble average on every statistical configuration of the field. Using the ergodic theorem, this mean can be identified with the spatial average of the same quantity. The average of any operator $X$ can be computed from the mean of its Fourier component $\tilde{X}(k)$
\begin{equation}
\label{avg_k}
\avg{X}:=\frac{1}{2\pi^2\sigma_0^2}\int_0^{+\infty} k^2 P(k)\tilde{X}(k) dk=\frac{\int_0^{+\infty} k^2 P(k)\tilde{X}(k) dk}{\int_0^{+\infty} k^2 P(k) dk}
\end{equation} 
Furthermore, we are interested in deriving the properties of the field subject to a set of linear constraints $\boldsymbol{C}=\left\{C_1,...,C_n\right\}$. Following \citet{Bertschinger1987}, each constraint $C_i$ can be written as
\begin{equation}
\label{constraint}
C_i[\delta] := \int W_i(\boldsymbol{x}_i-\boldsymbol{x})\delta(\boldsymbol{x})d\boldsymbol{x} = c_i
\end{equation}
where $W_i$ is the corresponding window function and $c_i$ its value. For example, constraining the value of the field to a certain $\delta_0$ at some point $\boldsymbol{x}_0$ leads to $W_i=\delta_D(\boldsymbol{x}-\boldsymbol{x}_0)$ and $c_i=\delta_0$. For $n$ constraints, the joint probability $d\mathcal{P}[\boldsymbol{C}]$ that the field satisfies these conditions reaches \citep{vandeWeygaert1996, Bertschinger1987}
\begin{equation}
\label{Pc}
d\mathcal{P}[\boldsymbol{C}] = \frac{1}{\sqrt{(2\pi)^n\det \boldsymbol{Q}}}\exp\left[-\frac{1}{2}\boldsymbol{C}^t.\boldsymbol{Q}^{-1}.\boldsymbol{C}\right]\prod_{i=1}^nd c_i
\end{equation}
where $\boldsymbol{Q}$ is the covariance matrix of the constraints defined through $\boldsymbol{Q}=\avg{\boldsymbol{C}^t.\boldsymbol{C}}$.

%-------------------------------
\subsection{The full joint Gaussian probability distribution}
%-------------------------------

In this section we derive the full joint Gaussian probability to find a CoSphere with a given set of parameters in GRF. Since these regions are build around extremum, we must include the peak conditions derived by \citet{BBKS}. A local extrema located at $\bx_0$ is defined by three conditions
\begin{align}
\label{d0}
&\delta(\boldsymbol{x}_0)= \nu\sigma_0\\
\label{dp0}
&\eta_i = \frac{\partial \delta(\boldsymbol{x}_0)}{\partial x_i}=0\\
\label{dpp0}
&\zeta_{ij}=\frac{\partial^2 \delta(\boldsymbol{x}_0)}{\partial x_i\partial x_j}
\end{align}
where \refeq{d0} gives the height of the peak in unit of the fluctuation level
\begin{equation}
\sigma_0=\left[\frac{1}{2\pi^2}\int_0^{+\infty} k^2 P(k) dk\right]^{1/2}
\end{equation}
whereas \refeq{dp0} imposes that the local gradient $\boldsymbol{\eta}$ vanishes (since we consider extrema). \refeq{dpp0} defines the Hessian matrix $\boldsymbol{\zeta}$ of the density profile around the peak. 

In addition to the peak condition, we must explicitly encode the compensation condition \refeq{R1}. This is achieved by adding the new constraints \paperIp
\begin{align}
\label{cons_R1}
C_{R_1}[\delta]&:=\int \Theta\left(R_1-|\boldsymbol{x}-\boldsymbol{x}_0|\right)\delta(\boldsymbol{x})d\boldsymbol{x}=\bar{\nu}\sigma_0=0\\
\label{cons_v1}
C_{\nu_1}[\delta]&:=\int \delta_D\left(R_1-|\boldsymbol{x}-\boldsymbol{x}_0|\right)\delta(\boldsymbol{x})d\boldsymbol{x}=\nu_1\sigma_0
\end{align}
where $\Theta$ is the Heaviside step function and $\delta_D$ is the usual Dirac delta.
\refeq{cons_R1} is the transposition of \refeq{R1} in the form \refeq{constraint}. The parameter $\bar{\nu}$ is defined by $\Delta(R_1)=\bar{\nu}\sigma_0$ and is set to $0$ by definition of the compensation radius $R_1$.
\refeq{cons_v1} defines the compensation density on the sphere of radius $R_1$ such that $\delta(R_1):=\delta_1=\nu_1\sigma_0$. 

\subsubsection{The full joint probability for spherically compensated peaks}

Without any assumption on the symmetry, CoSpheres in primordial field are described by $12$ independent scalars $(\nu,\bar{\nu},\nu_1,\eta_1,\eta_2,\eta_3,\zeta_{ij})$ with $i$ and $j$ running in $\{1, 2, 3\}$. The computation of the conditional probability \refeq{Pc} involves the correlation matrix $\boldsymbol{Q}$ between these $12$ constraints. The introduction of two new degree of freedom makes the computation of $\boldsymbol{Q}$ more complicated than for a standard unconstrained peak. However, following \citet{BBKS}, we can simplify $\boldsymbol{Q}$ by introducing the reduced variables linked to the local curvature of the profile around the peak
\begin{equation*}
\label{xyz}
x=-\frac{\zeta_{11}+\zeta_{22}+\zeta_{33}}{\sigma_0\sqrt{\avg{k^4}}},\quad 
y= -\frac{\zeta_{11}-\zeta_{33}}{2\sigma_0\sqrt{\avg{k^4}}}, \quad
z=-\frac{\zeta_{11}-2\zeta_{22}+\zeta_{33}}{2\sigma_0\sqrt{\avg{k^4}}}
\end{equation*}
where the various moments of $P(k)$ are given by
\begin{equation}
\avg{k^{2n}} := \frac{\sigma_n^2}{\sigma_0^2}=\frac{1}{2\pi^2\sigma_0^2}\int_0^{+\infty} k^{2 + 2n} P(k)dk
\end{equation}
$y$ and $z$ quantify the asymmetry of the profile around the peak whereas $x$ defines the local curvature. It is directly related to the spherical density profile by
\begin{equation}
\label{x_def}
\lim_{r\to 0}\frac{\partial^2\delta(r)}{\partial r^2}= -\frac{x}{3}\sigma_0\sqrt{\avg{k^4}}
\end{equation}
With these variables, $\boldsymbol{Q}$ reduces to a partitioned matrix where the only non diagonal terms are included in a $4\times 4$ sub-matrix $\tilde{\boldsymbol{Q}}$. This sub-matrix encodes the new correlations introduced by $R_1$ (or $\bar{\nu}$ equivalently) and $\nu_1$. In the $(\nu, \bar{\nu}, x, \nu_1)$ basis, it reaches
\begin{equation}
\label{sub}
\tilde{\boldsymbol{Q}}=\begin{pmatrix}
1 & \avg{W_1} & \frac{\avg{k^2}}{\sqrt{\avg{k^4}}} & \avg{J_1}\\
\avg{W_1} & \avg{W_1^2} & \frac{\avg{k^2W_1}}{\sqrt{\avg{k^4}}} & \avg{W_1J_1} \\
\frac{\avg{k^2}}{\sqrt{\avg{k^4}}} & \frac{\avg{k^2W_1}}{\sqrt{\avg{k^4}}} & 1 & \frac{\avg{k^2J_1}}{\sqrt{\avg{k^4}}}\\
\avg{J_1} & \avg{W_1J_1} & \frac{\avg{k^2J_1}}{\sqrt{\avg{k^4}}} & \avg{J_1^2}
\end{pmatrix}
\end{equation}
where we used the following notation for the spherical Bessel functions evaluated at $R_1$.
\begin{align}
W_1&:=3\frac{\sin(kR_1)-kR_1 \cos(kR_1)}{(kR_1)^3}\\
J_1&:=\frac{\sin(kR_1)}{kR_1}
\end{align}
We can now rewrite \refeq{Pc} as
\begin{equation}
d^{12}\mathcal{P}(\nu, \bar{\nu},x, \nu_1, y, z, \boldsymbol{\eta},\zeta_4,\zeta_5,\zeta_6)\propto \frac{1}{\sqrt{\det \boldsymbol{Q}}}\exp\left[-\frac{1}{2}\mathcal{F}\right]\mathcal{D}
\end{equation}
where the superscript $12$ indicates that this is $12$ dimensional quantity with the measure $\mathcal{D}=d\nu d\bar{\nu}dxd\nu_1dydz\prod_{i=4}^6d\zeta_{i}\prod_ld\eta_l$ with $\zeta_4=\zeta_{23}$, $\zeta_5=\zeta_{13}$ and $\zeta_6=\zeta_{12}$ \citep{BBKS}. 

We now neglect  the numerical factors which do not depend explicitly on $R_1$. The 2 form $\mathcal{F}$ reduces to
\begin{align}
\label{F}
\nonumber \mathcal{F}=& \frac{x^2C_x+\nu^2C_\nu+\nu_1^2C_{\nu_1}+2\left(x\nu C_{x\nu}+x\nu_1 C_{x\nu_1}+\nu_1\nu C_{\nu_1\nu}\right)}{\Sigma^2(R_1)}\\
&  + 15y^2 + 5z^2
\end{align}
where we have already imposed the condition  $\eta_i=0$ \refeql{dp0} and  $\bar{\nu}=0$ \refeql{cons_R1}. The  $C_\alpha$ functions (with $\alpha= 0, x, \nu, \nu_1, x\nu, x\nu_1, \nu_1\nu$) depend also on $R_1$. Their explicit form is given in \refapp{Calpha}. $\Sigma^2(R_1)$ takes the form
\begin{equation}
\label{Sigma2}
\Sigma^2(R_1)=C_0 + C_x+C_\nu+2\frac{\avg{k^2}}{\sqrt{\avg{k^4}}}C_{x\nu}
\end{equation}
Since we consider only spherical profiles, we marginalize over the asymmetry parameters $y$ and $z$. The integration of $d\mathcal{P}$ over $y$ and $z$, combined with the ordering condition $|\zeta_{11}|\geq|\zeta_{22}|\geq |\zeta_{33}|\geq 0$ then leads to the four dimensional joint probability for the spherically compensated cosmic regions 
\begin{equation}
d^4\mathcal{P}(\nu, x, \bar{\nu}, \nu_1) \propto \frac{f(x)}{\Sigma(R_1)}\exp\left[-\frac{\mathcal{L}(x,\nu,\nu_1)}{2}\right]d\nu dx d\bar{\nu}d\nu_1
\end{equation}
with \citep{BBKS}
\begin{align*}
f(x)=&\sqrt{\frac{2}{5\pi}}\left[e^{-\frac{5x^2}{2}}\left(-\frac{8}{5}+\frac{x^2}{2}\right)+e^{-\frac{5x^2}{8}}\left(\frac{8}{5}+\frac{31x^2}{4}\right)\right]\\
&+\frac{x^3-3x}{2}\left[\text{Erf}\left(x\sqrt{\frac{5}{8}}\right)+\text{Erf}\left(x\sqrt{\frac{5}{2}}\right)\right]
\end{align*}
This function is not modified here because it results from the integration over the y and z variables which are not correlated to $\nu_1$ nor $\bar{\nu}$. We define $\mathcal{L}$ as $\mathcal{L}:=\mathcal{F}-15y^2 - 5z^2$, \ie
\begin{align}
\nonumber\mathcal{L}(x,\nu,\nu_1,R_1)=&\frac{x^2C_x+\nu^2C_\nu+\nu_1^2C_{\nu_1}}{\Sigma^2(R_1)}\\
\label{L}
&+ 2\frac{x\nu C_{x\nu}+x\nu_1 C_{x\nu_1}+\nu_1\nu C_{\nu_1\nu}}{\Sigma^2(R_1)}
\end{align}
Note that $\mathcal{L}$ depends on $R_1$ through $\Sigma$ and the various $C_{\alpha}$ functions. When $R_1$ becomes very large, we recover the BBKS limit (see below \refsec{sec:R1_infty}) and $\mathcal{L}$ reduces to its expression as derived in \citet{BBKS}. Finally, we map $\bar{\nu}$ to the compensation radius as 
\begin{equation}
d\bar{\nu}=\abs{\frac{3\nu_1}{R_1}}dR_1
\end{equation}
and we get the full joint Gaussian probability distribution of CoSpheres
\begin{equation}
\label{Pnotfull}
d^4\mathcal{P}(\nu, x, \nu_1, R_1) \propto \frac{\abs{\nu_1}f(x)}{R_1\Sigma(R_1)}\exp\left[-\frac{\mathcal{L}(x,\nu,\nu_1,R_1)}{2}\right]d\nu dx d\nu_1dR_1
\end{equation}
where both $\Sigma(R_1)$ and $\mathcal{L}$ depend on $R_1$.

\subsubsection{The First Crossing Condition (FCC)}
\label{sec:FCC}

Our definition of $R_1$ \refeql{R1} implicitly assumes that $R_1$ is the first crossing radius such as $\Delta(R_1)=0$. However, neither \refeq{R1} nor the definition of $\nu_1$ insures it. For each $R_1$, there is a sub-domain for the shape parameters where the corresponding average mass contrast profile vanishes at some effective radius $\tilde{R}_1<R_1$. This is typically the case for central peaks with high curvature $x$. The true joint Gaussian probability must take this effect into account. In \paperI we show that the average mass contrast profile corresponding to a set of shape parameters $\nu$, $x$ and $\nu_1$ can be expressed as
\begin{equation}
\label{profile}
\Delta (r)=\sigma_0\left[\nu\Delta_\nu(r)+x\Delta_x(r)+\nu_1\Delta_{\nu_1}(r)\right]
\end{equation}
where each $\Delta_\alpha(r)$ function involves the compensation scale $R_1$ and the radius $r$. This set of shape parameters is safe if it satisfies 
\begin{equation}
\label{FCC}
\forall r\in[0, R_1[,\qquad 
\begin{cases}
\Delta (r)>0 \quad \text{if} \quad \nu >0 \\
\Delta (r)<0 \quad \text{if} \quad \nu <0 \\
\end{cases}
\end{equation}
This defines the safe domain $\mathcal{D}(R_1)$ for $\{\nu,x, \nu_1\}$ where the first radius where $\Delta(r)$ vanishes is $R_1$. If $\{\nu, x, \nu_1\}\notin \mathcal{D}(R_1)$ there exist an effective $\tilde{R}_1<R_1$ satisfying
\begin{equation}
\label{R_eff}
\nu\Delta_\nu(\tilde{R}_1)+x\Delta_x(\tilde{R}_1)+\nu_1\Delta_{\nu_1}(\tilde{R}_1)=0
\end{equation}
This effective compensation radius is associated with a compensation density $\tilde{\nu}_1$ defined as 
\begin{equation}
\label{v_eff}
\tilde{\nu}_1=\nu\delta_\nu(\tilde{R}_1)+x\delta_x(\tilde{R}_1)+\nu_1\delta_{\nu_1}(\tilde{R}_1)
\end{equation}
such that both $\tilde{R}_1$ and $\tilde{\nu_1}$ are functions of $\nu$, $x$, $\nu_1$ and $R_1$. The condition \refeq{FCC} defining the safe domain $\mathcal{D}(R_1)$ can be translated to a simple restriction on the curvature $x$
\begin{equation}
\label{xc}
\abs{x} < x_c(\nu, \nu_1, R_1)=\min\left(-\abs{\nu}\frac{\Delta_\nu(r)}{\Delta_x(r)} - \abs{\nu_1}\frac{\Delta_{\nu_1}(r)}{\Delta_x(r)},\quad\forall r < R_1\right)
\end{equation}
At fixed $R_1$, if $\abs{x}\geq x_c(\nu, \nu_1, R_1)$, then this set of parameters $\{R_1, \nu, x, \nu_1\}$ will contribute to $\{\tilde{R}_1, \nu, x, \tilde{\nu}_1\}$ where $\tilde{R}_1$ and $\tilde{\nu}_1$ are the effective parameters defined in \refeq{R_eff} and \refeq{v_eff}.

In other words, for each $R_1$, there is a fraction of its parameter's domain contributing to smaller $R_1^-<R_1$ while a fraction of larger compensation radii with $R_1^+>R_1$ also contribute to this $R_1$.
The full joint Gaussian probability can thus be formally decomposed in two parts
\begin{align}
\label{Pfull}
	d^4\mathcal{P}_{tot}&(\nu, x, \nu_1, R_1) \propto   \underbrace{\Theta\Big(x_c(\nu,\nu_1, R_1)-\abs{x}\Big)d^4\mathcal{P}(\nu, x, \nu_1, R_1)}_{\text{direct contribution}} \\
	\nonumber &+ \underbrace{\int_{R_1}^\infty dR_1^+ \int_{-\infty}^{0}d\nu_1^+d^4\mathcal{P}(\nu, x, \nu_1^+, R_1^+)\delta_D(\tilde{R}_1 - R_1)\delta_D(\tilde{\nu}_1 - \nu_1)}_{\text{contribution from higher compensation radii}}
\end{align}
The first term accounts for peaks satisfying the first crossing condition (FCC) while the second one is the contribution from peaks with higher compensation radii whose effective compensation radius $\tilde{R}_1$ equals $R_1$ and effective compensation density $\tilde{\nu}_1$ equals $\nu_1$. Note that naturally, this indirect contribution term provides $x$ satisfying \refeq{xc}. 

\subsubsection{The large scale limit and the BBKS distribution}
\label{sec:R1_infty}

In this section we focus on the very large scale behaviour of the full joint Gaussian probability distribution, \ie when $R_1\to +\infty$. 

For clarity, let us assume a power-law matter power spectrum smoothed with a Gaussian kernel, $P(k)\sim k^n\exp(-k^2R_f^2)$, where the power index $n$ is the effective power index \textit{at very small $k$}. In  the limit $R_1\to +\infty$,  the $C_\alpha$ parameters (see \refapp{Calpha}) reduce to simple power laws
\begin{align}
&C_x\propto \left(\frac{R_1}{R_f}\right)^{-5-n}\\
&\frac{C_\nu}{C_x} = 1, \quad \frac{C_{\nu_1}}{C_x}\propto\left(\frac{R_1}{R_f}\right)^2\\
&\frac{C_{xv}}{C_x}= -\gamma,\quad \frac{C_{\nu_1x}}{C_x}\propto \left(\frac{R_1}{R_f}\right)^{-1-n},\quad \frac{C_{\nu_1\nu}}{C_x}\propto \left(\frac{R_1}{R_f}\right)^{-1-n}\\
&\frac{C_0}{C_x}\to \gamma^2 - 1
\end{align}
where $\gamma:=\avg{k^2}/\sqrt{\avg{k^4}}$. Using these limits, the exponential term $\mathcal{L}$ simplifies to
\begin{equation}
\label{L_infty}
\mathcal{L}_\infty(x,\nu,\nu_1, R_1)\simeq\frac{x^2+ \nu^2-2\gamma x\nu}{1-\gamma^2} + 2\epsilon R_1^2\nu_1^2 + \order{R_1^{-1-n}}
\end{equation}
where $\epsilon$ is a positive parameter independent from $R_1$. We note two features for $\mathcal{L}$. The first concerns the $(x,\nu)$ dependence which takes the same exact form than in \citet{BBKS}. The second concerns the term involving $\nu_1$. It depends explicitly on $R_1$ and contributes to an overall $\exp(-\epsilon \nu_1^2R_1^2)$ factor in the full joint probability \refeq{Pfull}. For $R_1\to +\infty$, combined with the $\abs{\nu_1}$ pre-factor appearing in \refeq{Pfull}, it leads to a global $\delta_D(\nu_1)$ such that full joint probability distribution reduces to 
\begin{equation}
\label{P_infty}
d\mathcal{P}(\nu,x,\nu_1, R_1)\underset{R_1\to +\infty}{\to} d\mathcal{P}_{bbks}(x,\nu)\times \frac{\delta_D(\nu_1)}{R_1^{(1-n)/2}}d\nu_1dR_1
\end{equation}
where $d\mathcal{P}_{bbks}(\nu,x)$ is the standard joint probability peak derived in \citet{BBKS}. This limit shows that a central peak with a very large compensation radius is decorrelated from its cosmic environment. As a matter of fact, the full joint probability distribution \refeql{P_infty} is separated in two independent parts, one concerning the local extrema ($\nu$ and $x$ only) and the other involving $R_1$ and $\nu_1$, \ie concerning its large scale environment.

The FCC \refsecl{sec:FCC} condition constraining the value of $x$ \refeql{xc} deeply simplifies in this large radii regime where it reduces to 
\begin{equation}
\label{x_infinity}
\abs{x}\leq \frac{\nu}{\gamma}
\end{equation}
This means that the statistical properties of the central extrema involving $x$ and $\nu$ reduce, for very large compensation radius, to the standard "unconstrained" peak statistic of BBKS with smaller central curvature satisfying \refeq{x_infinity}.

%An other interesting aspect in this high $R_1$ limit concerns the explicit $R_1$ dependence. As the effective index $n$ of the matter power spectrum on very large scale ($k\to 0$) is $n=0.96$ \citep{Komatsu2009}, the full joint probability distribution tends to an almost scale invariant form with $d\mathcal{P}\sim R_1^{-0.02}dR_1$.

We emphasize that \refeq{P_infty} illustrates the progressive decoupling between the central peak and its environment. Large $R_1$ will be associated with universal central peaks whose local shape and properties are similar to BBKS.
%-----------------------------
\subsection{Statistical properties of the shape parameters in GRF}
\label{sec:parameters_statistics}
Large scale density and mass profiles of CoSpheres are described by four parameters within Gaussian random field \paperIp. These parameters are
\begin{enumerate}
	\item $\nu$ and $x$ (defined respectively in \refeq{d0} and \refeq{x_def}) characterizing the central extremum \citep{BBKS}
	\item the compensation radius $R_1$ itself \refeql{cons_R1} quantifying the size of the over/under massive sphere surrounding the central extremum
	\item the reduced compensation density $\nu_1$ \refeql{cons_v1} defined on the compensation sphere by $\delta(R_1)\equiv \delta_1=\nu_1\sigma_0$.
\end{enumerate}
This section is devoted to the study of the statistical properties of these shape parameters. Firstly, we compute the probability distribution of the compensation radius $R_1$ by marginalizing over the three other shape parameters. It provides the probability to find a $R_1$ whatever the central extrema and $\delta_1$. We then compute the marginalized conditional probability $d\mathcal{P}(X|R_1)$ for each shape parameter $X=\{\nu,x,\nu_1\}$ at fixed compensation radius. We use this distribution to deduce their conditional moments $\avg{X^n|R_1}$ within GRF. We finally discuss the physical properties of the mean average radial matter profile involving the mean value $\avg{X|R_1}$ for each shape parameter $X$.

In this whole section, we assume central maxima with $\nu>0$, $x >0$ and $\nu_1<0$. The treatment of the symmetric case (central under-density) is exactly symmetric and leads to the same results with the following substitutions $x\to -x$, $\nu \to -\nu$ and $\nu_1 \to -\nu_1$ and the appropriate integration domains.

\subsubsection{The compensation radius probability distribution}

Each extremum can be associated with a unique $R_1$ separating the collapsing and the expanding shells. The probability $dP(R_1)$ to find a local extremum with $R_1$ and whatever the other shape parameters is obtained by marginalizing \refeq{Pfull} over the three shape parameters $\nu$, $x$ and $\nu_1$, leading to
\begin{equation}
\label{R1_distribution}
\frac{dP(R_1)}{dR_1}=\alpha \int_{\nu_0}^{+\infty} \Jnm{0}{0}(\nu,R_1) d\nu
\end{equation}
with $\alpha$ a normalisation factor, insuring that $\int_0^{+\infty}dP(R_1)=1$
\begin{equation}
\alpha^{-1}=\int_0^{+\infty}\int_{\nu_0}^{+\infty} \Jnm{0}{0}(\nu,R_1) d\nu dR_1
\end{equation}
and the function
\begin{equation}
\label{J00}
\Jnm{0}{0}(\nu,R_1):=\int_{-\infty}^{0} \int_0^{x_c}\frac{d^4\mathcal{P}_{tot}(\nu, x, \nu_1, R_1)}{d\nu dR_1}
\end{equation}
where the integration on the local curvature $x$ is done over $[0, x_c(\nu,\nu_1)]$ due to the FCC condition \refsecl{sec:FCC}. Note that the integration over $\nu_1$ goes from -$\infty$ to $0$ since we consider here a central maxima. 

\begin{figure}
\centering \includegraphics[width=1.0\linewidth]{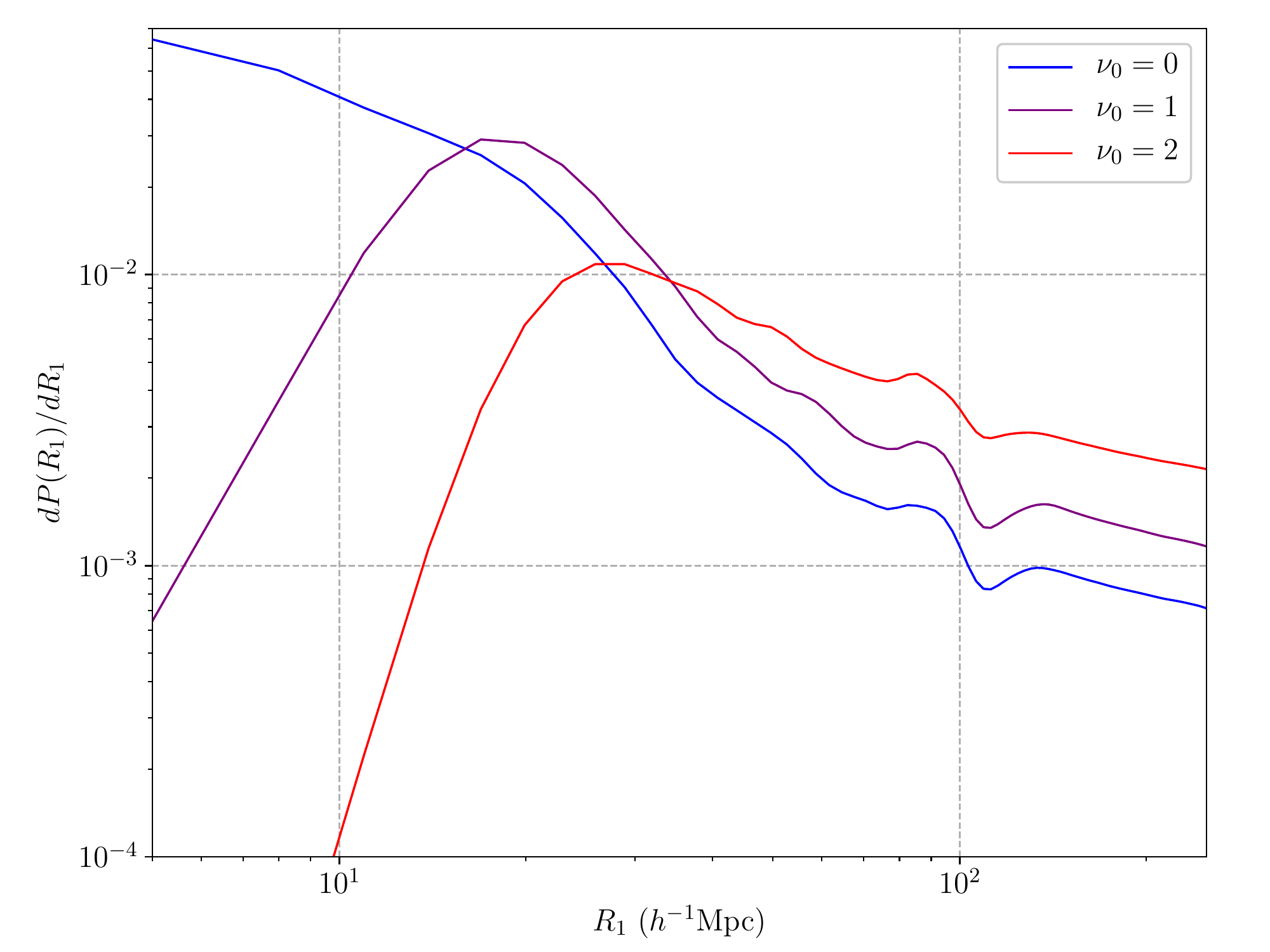}
    \caption{Probability distribution function for the compensation radius in GRF as computed in \refeq{R1_distribution}. Each curve corresponds to a different threshold $\nu_0$ which defines the minimal height of the central extremum, \ie $\abs{\nu}\geq \abs{\nu_0}$. Higher thresholds promote larger compensation radii. The most probable $R_1$ thus increases with $\nu_0$. For this figure, the $R_1$ pdf is normalized such that $\int_{0}^{300}dP(R_1)=1$ and the power spectrum has been smoothed with a Gaussian kernel on $R_g=5$ \Mpc}    
     \label{fig:PR1}
\end{figure}

On \refim{fig:PR1} we show this compensation radius probability $dP(R_1)$ for \LCDM cosmology in a Gaussian random field. We illustrate the effect of the central threshold $\nu_0$ defining the height of the central extrema $\abs{\nu}\geq \abs{\nu_0}$. Increasing the central threshold favors larger compensation radii. This seems natural since higher central peaks are more likely compensated on large regions than smaller ones. This figure also shows typical wiggles in this distribution around $R_1\sim 100$ \Mpc. This feature is probably related to the BAO. The enhanced correlation on this scale increases the probability to find CoSpheres compensated around this particular radius.

\subsubsection{The compensation density $\delta_1$}
\label{sec:d1}
The density contrast $\delta=\nu_1\sigma_0$ is measured on the compensation sphere at $r=R_1$. To get the joint probability for $\nu_1$ and $R_1$, we marginalize the full joint probability distribution \refeql{Pfull} over the central height $\nu$ and the curvature $x$
\begin{equation}
\label{Pv1R1}
\frac{d^2\mathcal{P}(\nu_1, R_1)}{d\nu_1dR_1}= \int_{\nu_0}^\infty \int_0^{x_c}\frac{d^4\mathcal{P}_{tot}(\nu, x, \nu_1, R_1)}{d\nu_1dR_1}
%\abs{\nu_1}\int_{\nu_0}^{+\infty}\int_0^{x_c}\frac{f(x)}{R_1\Sigma(R_1)}\exp\left(-\frac{\mathcal{L}(x,\nu,\nu_1,R_1)}{2}\right)dxd\nu
\end{equation} 
Note that $\nu$ is integrated from $\nu_0$ to $+\infty$ where $\nu_0$ is the lower threshold for the central height. The conditional probability $d\mathcal{P}(\nu_1|R_1)$ is deduced from Bayes theorem
\begin{equation}
\label{Pv1withR1}
\frac{d\mathcal{P}(\nu_1|R_1)}{d\nu_1}=\frac{\int_{\nu_0}^\infty \int_0^{x_c}d^4\frac{\mathcal{P}_{tot}(\nu, x, \nu_1, R_1)}{d\nu_1dR_1}}{\int_{\nu_0}^{+\infty}\Jnm{0}{0}(\nu, R_1)d\nu}
%\frac{\abs{\nu_1}\int_{\nu_0}^{+\infty} d\nu \int_0^{x_c}f(x)dx\exp\left[-\frac{\mathcal{L}}{2}\right]}{\int_{\nu_0}^{+\infty}\Jnm{0}{0}(\nu, R_1)d\nu}
\end{equation}
which describes the probability to get a compensated region with $\nu_1$ given $R_1$ normalized such that $\int_{-\infty}^{0}d\mathcal{P}(\nu_1|R_1)=1$.

\begin{figure}
	\centering \includegraphics[width=1.0\linewidth]{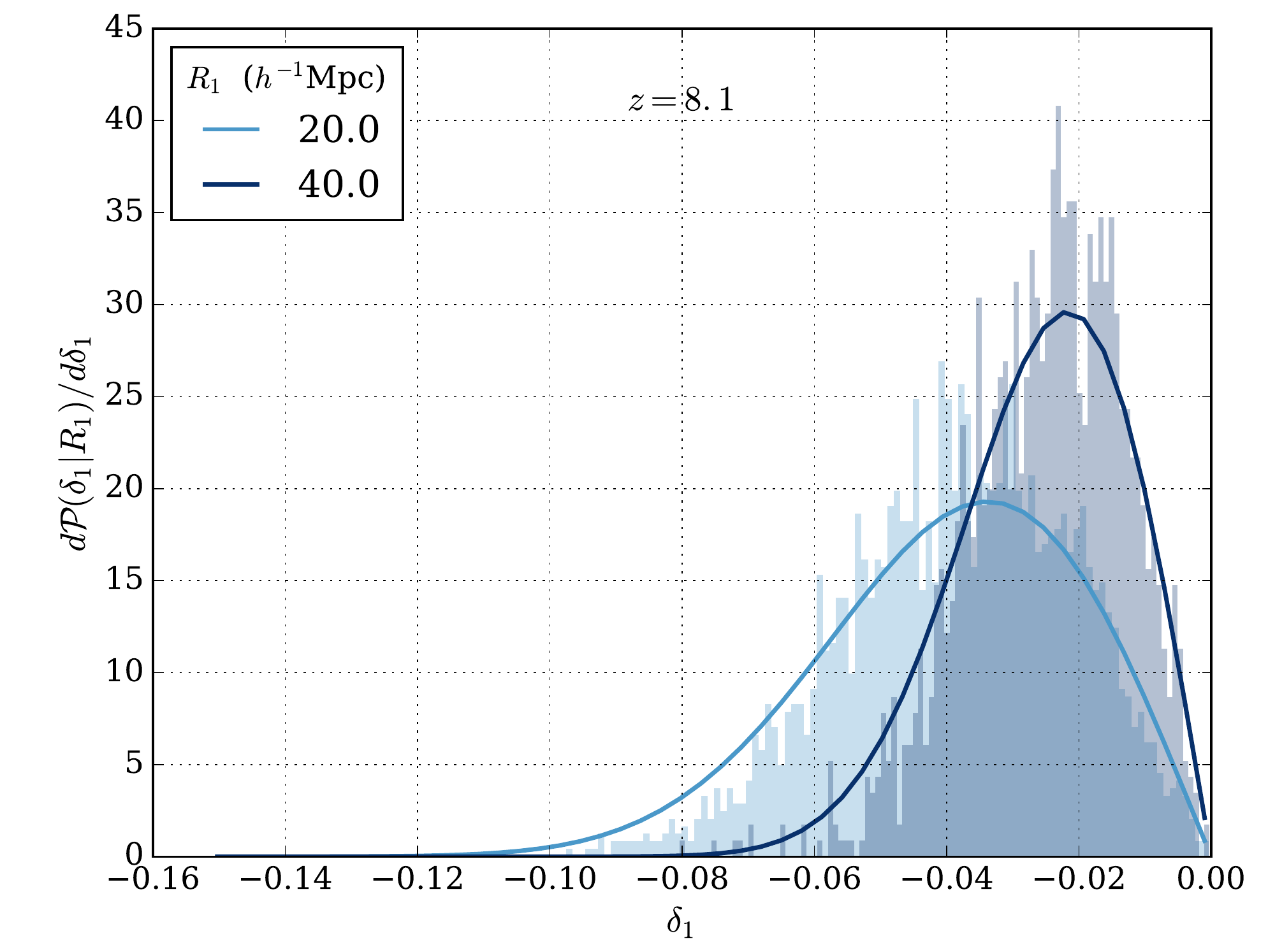}
	\caption{Probability density function $d\mathcal{P}(\delta_1|R_1)/d\delta_1$ computed from \refeq{Pv1withR1} in the Gaussian field at $z=8.1$. Curves are the theoretical expectations for the \LCDM model while the shaded regions are the measured distributions in the reference simulation for two different compensation radius in the case of a central maximum (thus negative values of $\delta_1$).}
		\label{fig:distribution_d1_init}
\end{figure}

On \refim{fig:distribution_d1_init} we plot the distribution of $\delta_1$ in a Gaussian random field with a comparison to numerical simulation, illustrating the excellent agreement between the theoretical expectation and the numerical results. As an illustration, if we neglect the dependence of $x_c$ in term of $\nu_1$ and the second term in \refeq{Pfull}, $\nu_1$ follows a distribution of the form
\begin{equation}
\label{Pv1R1_approx}
\frac{d\mathcal{P}(\nu_1|R_1)}{d\nu_1} \propto \abs{\nu_1}\exp\left[-\frac{(\nu_1-\bar{\nu}_1)^2}{2\sigma^2}\right]
\end{equation}
where  $\bar{\nu}_1$ and $\sigma$ are respectively the mean and dispersion value of the $\nu_1$ distribution and are both functions of $R_1$.

From \refeq{Pv1withR1} we compute the  moments of $\nu_1$ given $R_1$, defined by 
\begin{equation}
\label{mean_v1_n}
\avg{\nu_1^n| R_1}=\frac{\int_{\nu_0}^{+\infty}\Jnm{n}{0}(\nu, R_1)d\nu}{\int_{\nu_0}^{+\infty}\Jnm{0}{0}(\nu)d\nu}
\end{equation}
where $\Jnm{n}{m}$ generalizes the function defined in \refeq{J00} as
\begin{align}
\nonumber \Jnm{n}{m}(\nu, R_1):=\int_{-\infty}^{0} \nu_1^n\int_0^{x_c}x^m\frac{d^4\mathcal{P}_{tot}(\nu, x, \nu_1, R_1)}{d\nu dR_1}
\end{align}
For $n=1$ we get the average value of $\nu_1$ given $R_1$
\begin{equation}
\label{avgV1}
\avg{\nu_1|R_1}=\frac{\int_{\nu_0}^{+\infty}\Jnm{1}{0}(\nu,R_1)d\nu}{\int_{\nu_0}^{+\infty}\Jnm{0}{0}(\nu,R_1)d\nu}
\end{equation}
On \refim{fig:parameters}, we plot $\avg{\nu_1|R_1}$ as a function of the compensation radius $R_1$ in a Gaussian random field. 
$\avg{\nu_1|R_1}$ (red curve) admit a maximum for small compensation radius (here $R_1\sim 5$ \Mpc as we used a Gaussian smoothing scale $R_f=2$ \Mpc for the matter power spectrum) and slowly converges to $0$. 

\subsubsection{The heigh of the central peak $\nu$}
\label{sec:v_stat}
The conditional probability distribution of the heigh $\nu$ of the central extremum given $R_1$ is obtained by integrating the full joint probability \refeql{Pfull} over $x$ and $\nu_1$,
\begin{equation}
\frac{d\mathcal{P}(\nu|R_1)}{d\nu}= \frac{\Jnm{0}{0}(\nu,R_1)}{\int_{\nu_0}^{+\infty}\Jnm{0}{0}(\nu, R_1)d\nu}
\end{equation}
we deduce the moments of $\nu$ constrained by its cosmic environment i.e. for a given compensation radius
\begin{equation}
\label{mean_v_n}
\avg{\nu^n|R_1}=\frac{\int_{\nu_0}^{+\infty} \nu^n\Jnm{0}{0}(\nu,R_1)d\nu}{\int_{\nu_0}^{+\infty}\Jnm{0}{0}(\nu,R_1)d\nu}
\end{equation}
and in particular the average value for $\nu$ obtained for $n=1$
\begin{equation}
\label{mean_v}
\avg{\nu|R_1}=\frac{\int_{\nu_0}^{+\infty} \nu\Jnm{0}{0}(\nu)d\nu}{\int_{\nu_0}^{+\infty}\Jnm{0}{0}(\nu)d\nu}
\end{equation}
As it can be seen in \refim{fig:parameters}, $\avg{\nu|R_1}$ strongly depends on $R_1$ for small compensation radius while it progressively tends to its asymptotic value. As shown in \refsec{sec:R1_infty}, it converges to the standard value $\avg{\nu}$ computed by \citet{BBKS}. 

Small inhomogeneous regions (small $R_1$) are associated with lower central extremum, describing smoothed inhomogeneities while higher extremum (or deeper voids) are more likely to sit in larger over massive (resp. under massive) regions. As discussed in \refsec{sec:R1_infty}, the convergence toward the standard BBKS case illustrates the progressive decorrelation between the central peak and its large scale cosmic environment.

\subsubsection{The curvature distribution $x$}
\label{sec:x_stat}
\begin{figure}
	\centering \includegraphics[width=1.0\linewidth]{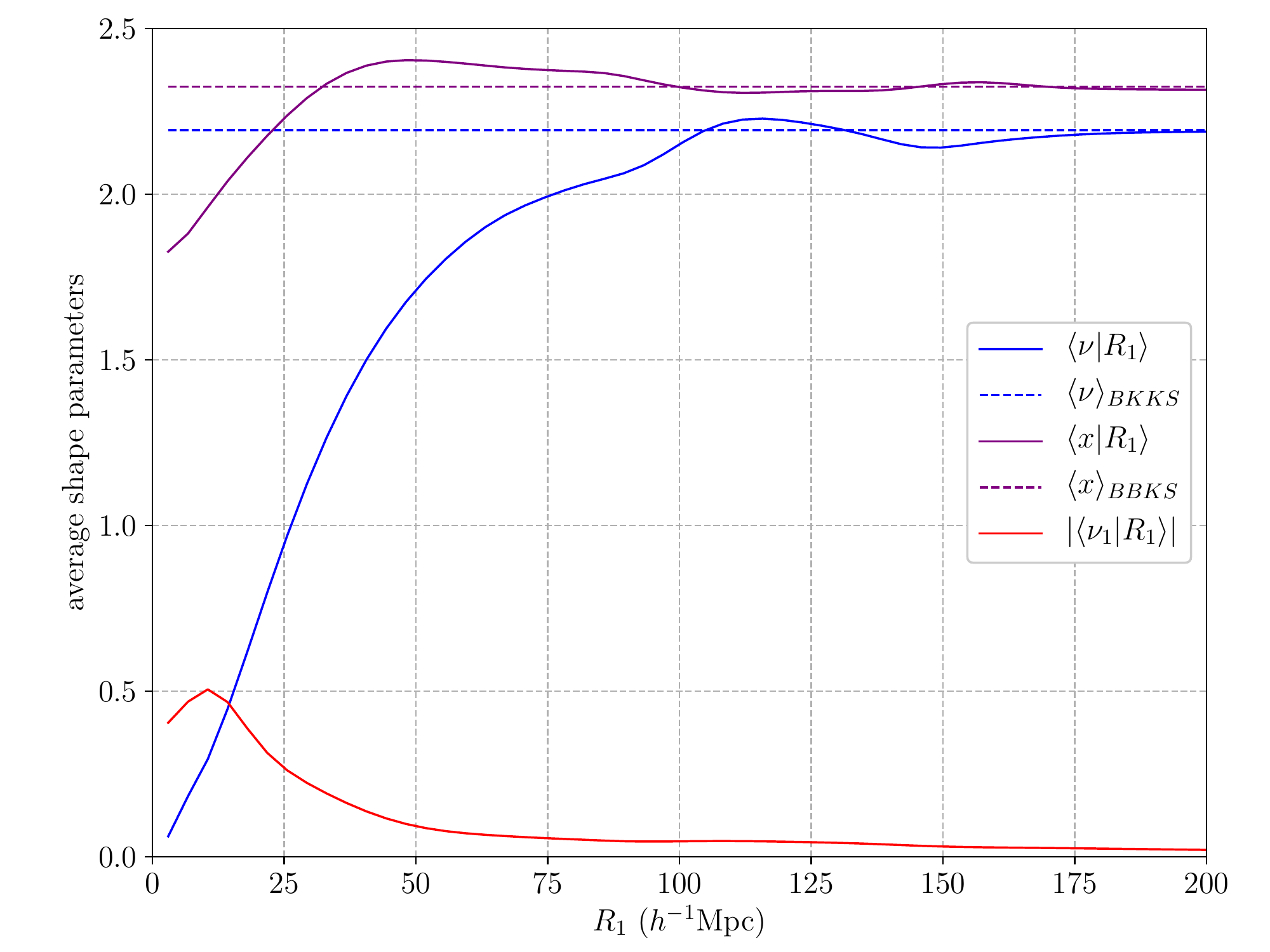}
	\caption{Mean expected values $\avg{X|R_1}$ for the shape parameters $X=\{\nu,x, \nu_1\}$ computed from \refeq{avgV1}, \refeq{mean_v} and \refeq{mean_x} as a function of the compensation radius. For illustration, the \LCDM matter power spectrum has been smoothed on a Gaussian scale $R_f=10$ \Mpc. The dashed lines are the expected values from \citet{BBKS} with the condition \refeq{x_infinity} and are recovered for $R_1\to\infty$ as shown in \refsec{sec:R1_infty}. Note that $\avg{\nu_1|R_1}\to 0$ in the $R_1\to \infty$ limit.}
		\label{fig:parameters}
\end{figure}

Finally we evaluate the statistical properties of the local curvature $x$ around a central extremum. Following the same development as before, we derive the various moments
\begin{equation}
\label{mean_x_n}
\avg{x^n|R_1} = \frac{\int_{\nu_0}^{+\infty} \Jnm{0}{n}(\nu,R_1) d\nu}{\int_{\nu_0}^{+\infty} \Jnm{0}{0}(\nu,R_1) d\nu}
\end{equation}
with the average of $x$ given by 
\begin{equation}
\label{mean_x}
\avg{x|R_1} = \frac{\int_{\nu_0}^{+\infty} \Jnm{0}{1}(\nu,R_1) d\nu}{\int_{\nu_0}^{+\infty} \Jnm{0}{0}(\nu,R_1) d\nu}
\end{equation}
We show on \refim{fig:parameters} the behavior of $\avg{x|R_1}$ as a function of $R_1$. For large compensation radii, it converges to its modified BBKS value (see \refsec{sec:R1_infty}) and remains almost constant for a wide range of $R_1$. Again we observe on BAO scale some wiggles for  $\avg{\nu|R_1}$  and $\avg{x|R_1}$ relating the peaks parameters and the compensation radius.

%---------------------------------
\subsection{The mean average profile with a given compensation radius $R_1$ in GRF}
\label{sec:averaged_profiles}

\subsubsection{The profile at fixed $R_1$}

\begin{figure}
	\centering \includegraphics[width=1.0\linewidth]{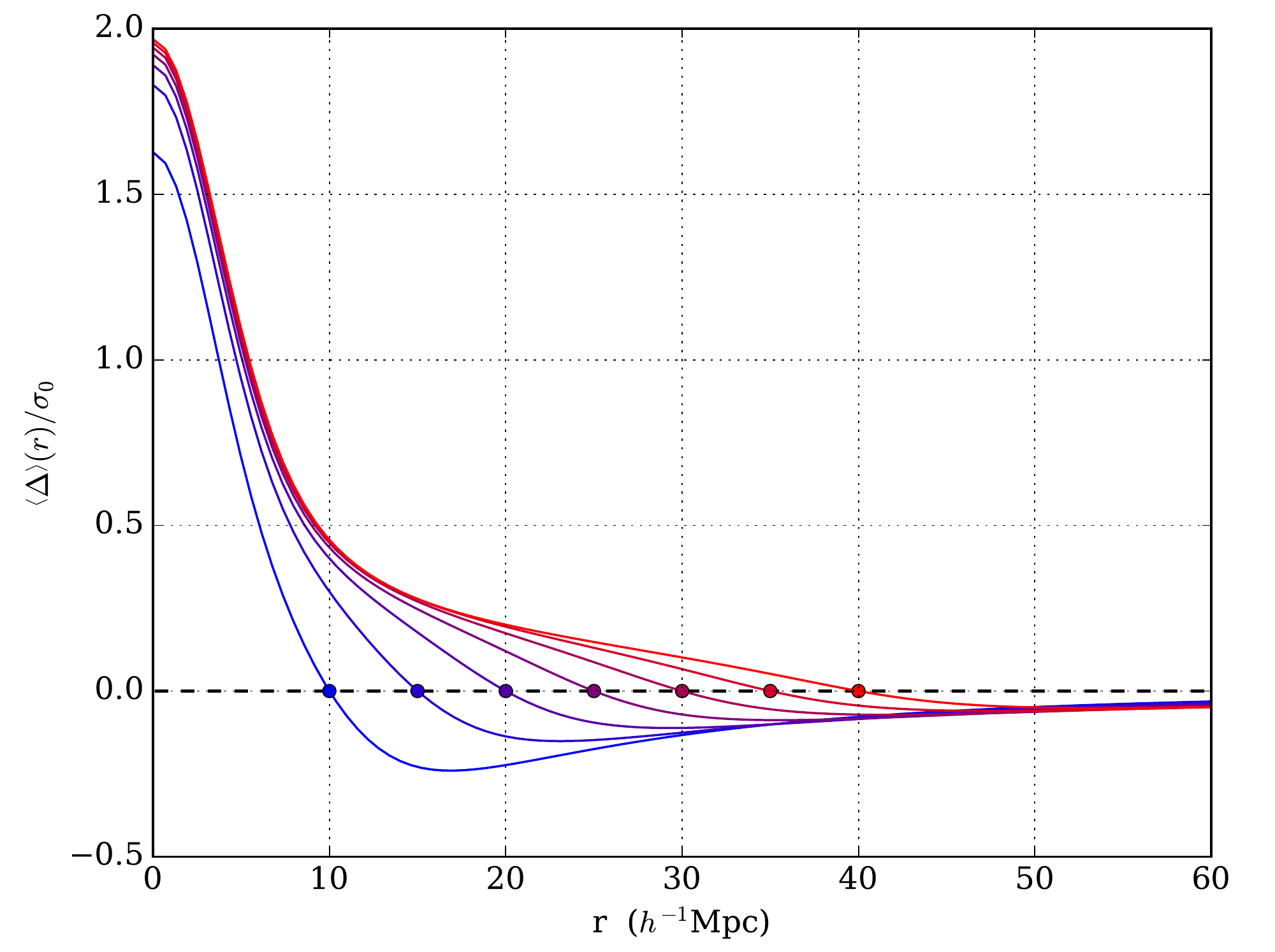}
	\caption{Mean average mass contrast profiles normalised to the fluctuation $rms$ $\sigma_0$ for various compensation radius $R_1$ (from 10 \Mpc to 40 \Mpc) in a GRF \refeql{mean_averaged_profile_mass}. The elbow appearing beyond $r\sim 10$ \Mpc for profiles with $R_1 > 30$ \Mpc is not due to any dynamical feature, it is already present in the Gaussian Random Field and results from the compensation constraint. It illustrates that while the compensation radius $R_1$ increases, the central extrema is progressively isolated from its surrounding cosmic environment and its shape tends to the universal BBKS profile.}
	\label{fig:mean_profiles}
\end{figure}

In the primordial Gaussian field, average profiles of CoSpheres are determined by four independent - but correlated - scalars; $\nu$, $x$, $\nu_1$ and $R_1$. At fixed compensation radius $R_1$, the other shape parameters $X=\{x, \nu, \nu_1\}$ can be considered as stochastic variables with constrained probabilistic distributions $d\mathcal{P}(X|R_1)$ as computed in the previous sections. Since the average density and mass contrast profiles are linear in the shape parameters (see \paperI and \refeq{mean_averaged_profile}), one can define the mean average profile at a fixed compensation radius as the profile whose shape parameters are averaged over their distribution, thus reaching

\begin{equation}
\label{mean_averaged_profile}
\frac{\overline{\avg{\delta}}(r)}{\sigma_0}=\avg{\nu|R_1}\delta_\nu(r)+\avg{x|R_1}\delta_x(r)+\avg{\nu_1|R_1}\delta_{\nu_1}(r)
\end{equation}
where brackets mean an average on stochastic realization of the field and bar means an average over the possible values for the free shape parameters. The mass contrast profile reaches \paperIp
\begin{equation}
\label{mean_averaged_profile_mass}
\frac{\overline{\avg{\Delta}}(r)}{\sigma_0}=\avg{\nu|R_1}\Delta_\nu(r)+\avg{x|R_1}\Delta_x(r)+\avg{\nu_1|R_1}\Delta_{\nu_1}(r)
\end{equation}
This profile describes the spherically compensated matter distribution resulting from stacking every possible realization at fixed $R_1$. On \refim{fig:mean_profiles} we show the mass contrast profiles $\overline{\avg{\Delta}}$ for various compensation radii in \LCDM cosmology. We retrieve the various properties of CoSpheres described before: (i) smaller central maxima (low $\nu$) are associated with narrow compensation radius with a deep compensation density $\delta_1$, (ii) higher central maxima (high $\nu$) are located in larger over massive regions with a high $R_1$ and smoother density contrast $\delta_1$ and (iii) when $R_1$ increases, central peaks become undistinguishable on small scales ($r \ll R_1$) and tend to the standard BBKS profiles. In other words, for large $R_1$, different environments with various compensation radii can be associated with very similar central profiles.

The whole of the previous discussion can be directly transposed to the symmetric case of a central minima, seeding cosmic void.

\subsubsection{On the characteristic elbow}

One particular feature of the mean average profile, besides the fact that they are fully determined by one single parameter $R_1$, is the existence of a characteristic elbow. This bend appears around $r\sim 10$ \Mpc in \refim{fig:mean_profiles} but it also shows up in numerical profiles as can be seen in \refim{fig:profile_D_void_today}. This elbow is a result of the progressive decorrelation between the central peak and its surrounding environment as discussed in \refsec{sec:R1_infty}. 

While $R_1$ increases, the central extremum tends to an universal shape as expected from BBKS. This elbow appears as the transition between small "BBKS" scales and larger ones involved with the compensation property. This characteristic does not appears in standard void profiles when build from their effective size $R_{eff}$ as in \cite{Hamaus2014}. This is likely due to the fact that voids with the same $R_{eff}$ may have very different compensation radii. Stacking together profiles with the same $R_{eff}$ may erase this feature. On the other hand, this elbow does not appears in evolved profiles build from central over densities as in \refim{fig:profile_D_today} despite existing in the primordial Universe (see \refim{fig:mean_profiles}). This vanishing follows from the non linear gravitational evolution of these profiles, altering their shape on small scales.

%%%%%%%%%%%%%%%%%%
%% %   SECTION 3    %%%% %%
%%%%%%%%%%%%%%%%%%
\FloatBarrier
\section{Non linear gravitational evolution of CoSphere in \LCDM cosmology}
\label{dynamics}

In the previous section we discussed the statistical properties of the shape parameters of CoSpheres within the primordial GRF. These results stand under the Gaussian assumption which can be safely assumed at high redshift. In this section we study the dynamical evolution of these quantities during the non linear collapse of the matter field. As shown in \paperI, the adapted formalism for the gravitational collapse of these regions is the Lagrangian spherical collapse model \citep{Padmanabhan1993,Peacock1998}. It describes the Lagrangian evolution of concentric shells without shell-crossing or caustics formation. 

%------------------
\subsection{Spherical Lagrangian collapse in $\Lambda$CDM cosmology}
\label{sec:dynamic}

We recall the dynamical equations for the Lagrangian collapse suited for our study. In the following, a Greek letter $\Chi$ will denote a \textit{comoving} quantity while a Latin character $r$ designates a \textit{physical} length. These quantities are related by $r=\Chi\times a$ with $a$ the homogeneous scale factor normalized as $a(t_0)=1$ today. We also denote every initial quantity by the ''$i$'' label, $e.g.$ $\Chi_i$ is the initial comoving position of one shell. Initial conditions are taken deep in the matter dominated era where The Gaussian assumption for $\delta(\bx)$ stands. We define the dimensionless Lagrangian displacement for each shell
\begin{equation}
\label{normalisation}
\varsc(\Chi_i,t)=\frac{\Chi(t)}{\Chi_i}\quad
\end{equation}
with $\Chi(t)$ the comoving radius of the shell at some time $t$. The mass conservation in the absence of shell crossing leads to the relation
\begin{equation}
\label{fnoSC}
\frac{1+\Delta}{1+\Delta_i}=\varsc^{-3}
\end{equation}
where $\Delta_i$ is the initial mass contrast for this shell, \ie $\Delta_i=\Delta(\Chi_i)$ and 
$\Delta$ its evolved mass contrast. In order to simplify the dynamical equation, we introduce the affine parameter $\vart$ defined through 
\begin{equation}
\label{tildeaplha}
\frac{d \vart}{d\log(a)}:=\sqrt{\frac{\Omega_m}{2}}
\end{equation}
which can be integrated to give $\vart(a)$ in the $\Lambda$CDM model, with the definition $\vart(a_i)=0$
\begin{equation}
\label{theta}
\vart(a)=\frac{\sqrt{2}}{3}\left[\arctanh\left(\Omega_{m,i}^{-1/2}\right)-\arctanh\left(\Omega_m^{-1/2}\right)\right]
\end{equation}
With this new parametrization, the equation of motion driving the evolution of each individual shell reaches \paperIp
\begin{equation}
\label{dynamic}
\frac{\partial^2 \varsc}{\partial\vart^2}+\frac{1}{\sqrt{2\Omega_m}}\frac{\partial\varsc}{\partial\vart} = \varsc-\frac{1+\Delta_i}{\varsc^2}
\end{equation}
To close our system we need to specify the initial conditions at $\vart=0$. They are fixed by assuming that the dynamics follows the Zel'dovich evolution at very high redshift, leading to \paperIp
\begin{equation}
\label{ini1}
\begin{cases}
\varsc(t_i)&=1\\
\frac{\partial\varsc}{\partial \vart}(t_i)&=-\sqrt{\frac{2}{\Omega_{m,i}}}\frac{\Delta_i}{3}f(t_i)
\end{cases}
\end{equation}
where $f$ is the linear growth rate \citep{Peebles1980} and $f(t_i)$ is evaluated at the initial time defined by $\vart=0 \Leftrightarrow t=t_i$. \refeq{dynamic} is valid for any cosmology with a quintessence field sourcing dark energy and possibly a time varying e.o.s parameter $w$. The affine parameter $\vart$ is then still defined by \refeq{tildeaplha} but \refeq{theta} is no longer true \citep{paper3}. We extend also \refeq{dynamic} for theories beyond GR in \citet{paper4}.

%-------------------------
\subsection{Dynamical evolution of the compensation radius probability distribution}
\label{SizeEvolution}

The particular scale $R_1$ is by definition conserved in comoving coordinates, $i.e.$ $R_1(t)\propto a(t)$. In other terms, since the mean density enclosed in the sphere of radius $R_1$ equals the background density, this scale evolves as the scale factor of the Universe. Since $R_1$ is conserved, its probability distribution must also be conserved during the gravitational evolution. In principle, merging or creation of local extrema could modify this probability distribution. However, such effects are expected to occur on small scales, and since we consider sufficiently large value for $R_1$ ($R_1 \gtrsim 5-7$ \Mpc), the probability distribution $dP(R_1)$ will not be affected. 

\begin{figure}
	\includegraphics[width=1.0\linewidth]{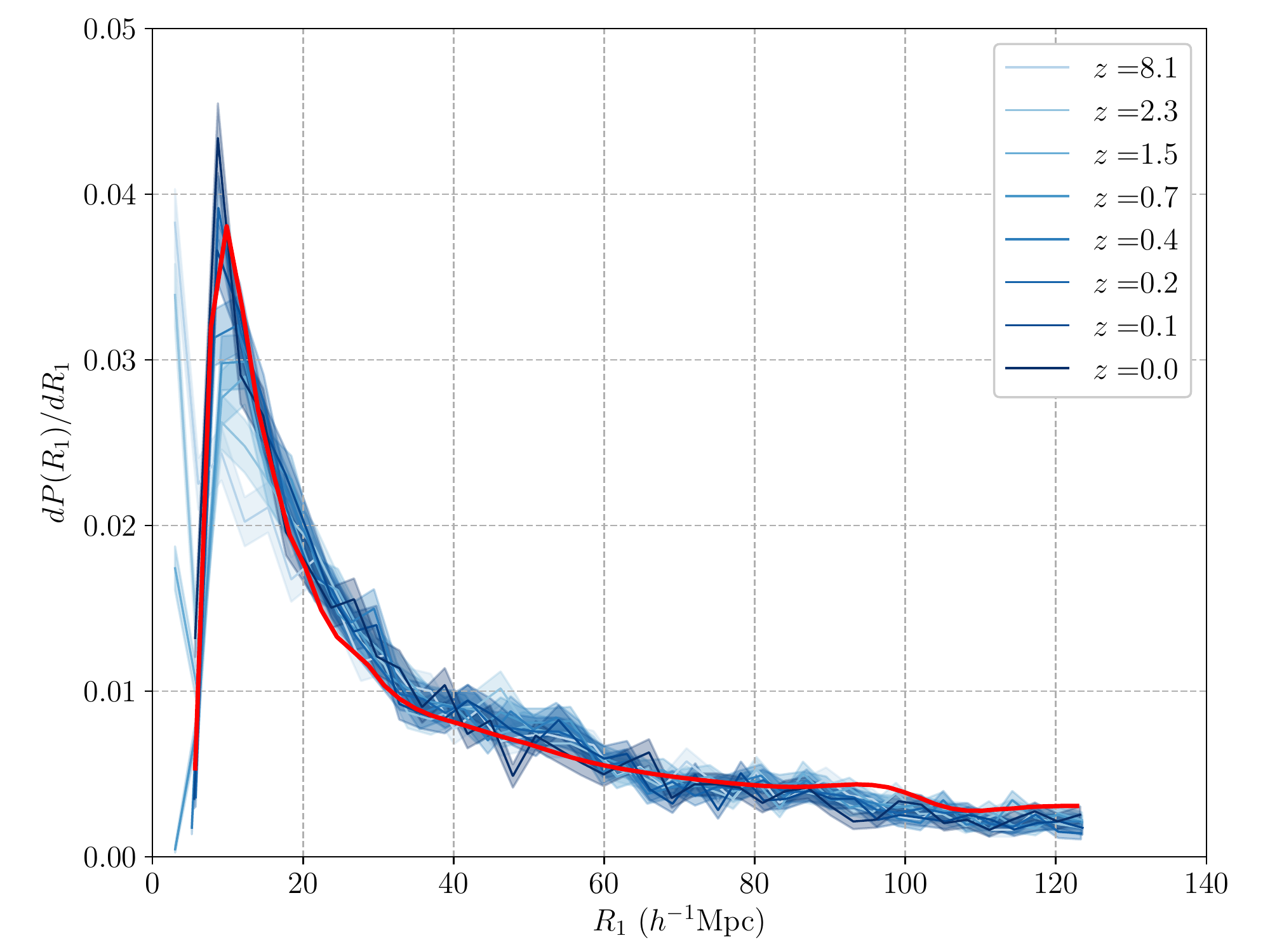}
	\caption{The compensation radius probability distribution $dP(R_1)$ for $\Lambda$CDM cosmology computed from $10000$ profiles build around haloes at various redshift ($z=8$ to $z=0$) in the reference simulation (blue lines). This figure has been obtained from $10000$ halos of mass $M_h= 3.0 \pm 0.075\times 10^{13}$ \Msun. The shaded region around each curve is the Poisson noise computed in radial bins of size $dR_1=3.5$ \Mpc. The red line is the initial Gaussian distribution given by \refeq{R1_distribution}. The conservation of $R_1$ insures the conservation of its probability distribution. The height of the central threshold $\nu_0$ has been chosen in agreement with the halo masses ($3.0 \times 10^{13}$ \Msun) used to construct the compensated regions.}
	\label{fig:PR1_z0_theory}
\end{figure}

On \refim{fig:PR1_z0_theory} we show the measures of its pdf $dP(R_1)/dR_1$ at various redshifts from $z=8$ to $z=0$ in the numerical simulation. We also show the theoretical expectation from \refeq{R1_distribution} computed within GRF. This figure illustrates two points. Firstly, the compensation radius pdf does not evolve during the cosmic evolution excepted on very small scales ($R_1\leq 5$ \Mpc) where our reconstruction procedure may be inaccurate \refsecl{section1}. On larger scales however, neither the shape nor the amplitude are affected, confirming that this distribution is conserved during cosmic history.

On the other hand, the GRF expectation \refeql{R1_distribution} fits the measured distribution with a very good agreement. This distribution thus appears as a good way to probe the early universe. However, since the initial power spectrum $P(k)$ is independent from the $e.o.s$ parameter for DE $w$, this probability distribution does not probe $w$ neither $\sigma_8$, the amplitude of the power spectrum, but may probe $\Omega_m$ and the various quantities describing the primordial Universe as the scalar index $n_s$ on very large scales. These cosmological dependences are discussed in \citet{paper3}.

Note however that the wiggles predicted by theoretical prediction around $R_1\simeq 100$ \Mpc do not appears in numerical data. This may be due to the finite volume of our simulation ($L_{box}=2592$ \Mpc) and the fact that on such scale, we are dominated by the cosmic variance \citep{Rasera2013}.

%------------------------
\subsection{The evolution of the compensation density $\delta_1$}
\label{sec:d1_dynamics}

\begin{figure}
\centering \includegraphics[width=1.0\linewidth]{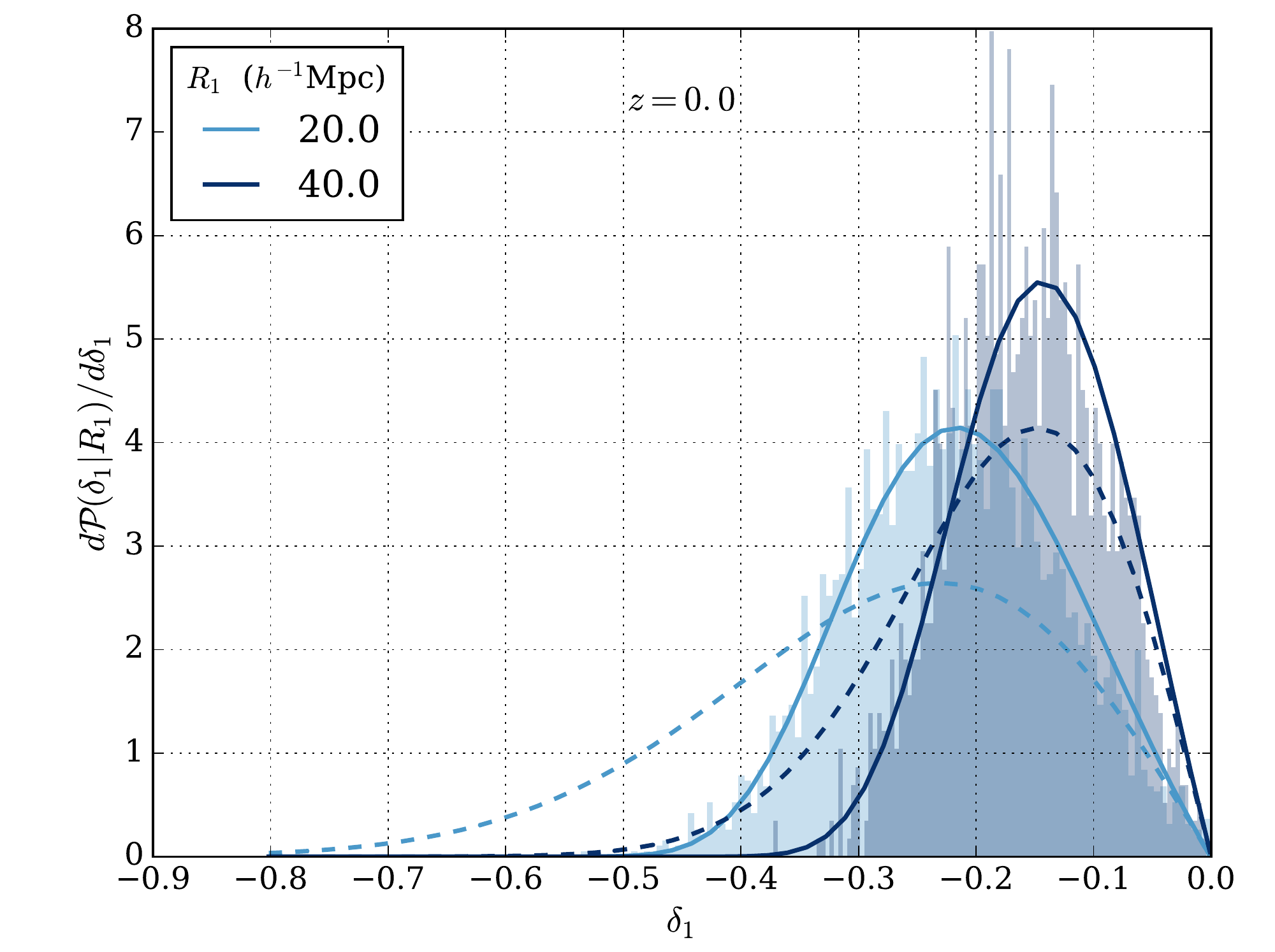}
    \caption{Evolved probability density function $dP(\delta_1|R_1)/d\delta_1$ at $z=0$ in the \LCDM cosmology from haloes. The full line curves corresponds to the exact evolution given by \refeq{Pd1_t}. The shaded regions are the measured distributions for two different compensation radius in the reference simulation. The dashed curves are the Gaussian prediction, \ie the linear evolution of the primordial distribution. This figure illustrates the non linearity of the local gravitational process but also the possibility to reproduce the evolved distribution from the exact collapse.}
    \label{fig:distribution_d1_today}
\end{figure}

$\delta_1$ is defined \textit{on the sphere} of radius $R_1$. It is a fundamental Eulerian quantity and its probability distribution can be computed analytically in the primordial GRF \refsecl{sec:d1}. $\delta_1$ is  directly measurable from the matter profile. In \paperI we showed that during the non linear evolution, it follows a simple dynamics, corresponding to a one-dimensional Zel'dovich collapse
\begin{equation}
\label{deltat}
\delta_1^t=\delta_1\frac{\tilde{D}(t)}{1 - \delta_1\left(\tilde{D}(t)-1\right)}
\end{equation}
where $\tilde{D}(t)$ is the normalized linear growth factor defined by $\tilde{D}(t)=D(t)/D(t_i)$ and $\delta_1^t=\delta_1(t)$ while $\delta_1=\delta_1(t_i)$ is its corresponding value in GRF. \refeq{deltat} only holds at the particular point  $r=R_1$ and cannot be extended to other arbitrary scale where Zel'dovich dynamics is, at best, an approximation. There is a bijective mapping between $\delta_1^t$ and $\delta_1$ insuring that \refeq{deltat} can be inverted 
\begin{equation}
\label{delta0}
\delta_1 = \frac{\delta_1^t}{\tilde{D}(t) + \delta_1^t\left(\tilde{D}(t)-1\right)}
\end{equation}
The computation of the non linearly evolved conditional probability distribution $dP(\delta_1^t|R_1)$ can be computed under the assumption that $R_1$ and the joint probability of $\delta_1$ and $R_1$ are both conserved during evolution. Since since $\delta_1$ and $\delta_1^t$ are connected with a one-to-one relation we have
\begin{equation}
dP(\nu_1^t|R_1)=dP(\nu_1|R_1)
\end{equation}
with $\nu_1^t=\delta_1^t/\sigma_0$ and $\nu_1=\delta_1/\sigma_0$ where $\sigma_0$ is computed in the primordial GRF only (and is a constant). Using \refeq{deltat}, we get the conditional probability distribution at any time
\begin{equation}
\label{Pd1_t}
dP(\delta_1^t|R_1)=\frac{\tilde{D}(t)}{\left[\tilde{D}(t)+\delta_1^t(\tilde{D}(t)-1)\right]^2}\left.\frac{dP(\nu_1|R_1)}{d\nu_1}\right|_{\nu_1=\frac{\delta_1}{\sigma_0}}\frac{d\delta_1^t}{\sigma_0}
\end{equation}
Where $\delta_1$ and $\delta_1^t$ are linked by \refeq{delta0}. In \refim{fig:distribution_d1_today} we show the distribution $dP(\delta_1^t|R_1)$ measured today in the reference simulation. Each colour corresponds to a compensation radius (here $20$ in red and $40$ \Mpc in blue). In each case we show the non linear prediction \refeq{Pd1_t} in full line together with the linear evolution in dashed lines. The full spherical prediction reproduces the measured distribution with a high accuracy whereas linear prediction predicts larger values of $\delta_1$ today, especially for small compensation radii. It is interesting to note that the linear prediction also fails on large scales usually considered as "linear", e.g. $R_1=40$ \Mpc. This difference come from the fact that despite being on "linear" scales, this distribution probes high density contrasts ($\delta_1$ around $-0.5$) which are in the non linear dynamical regime.

From \refeq{Pd1_t}, we can derive the average moments\footnote{despite being a mute parameter, we prefer to keep the notation $\delta_1^t$ in the integral to highlight the fact that this average is evaluated at any time and not only in the initial conditions} of $\delta_1^t$
\begin{equation}
\label{d1t_0}
\avg{\delta_1^n|R_1}(t)=\int_{-1}^0\left(\delta_1^t\right)^ndP(\delta_1^t|R_1)
\end{equation}
where the integration is done over $\delta_1^t$. Mapping $\delta_1^t$ to its corresponding value in the initial conditions $\delta_1$ leads to
\begin{equation}
\label{dt1}
\avg{\delta_1^n|R_1}(t)=\int_{-1}^0\left(\frac{\tilde{D}(t)\delta_1}{1- \delta_1(\tilde{D}(t)-1)}\right)^ndP(\delta_1|R_1)
\end{equation}

\begin{figure}
\centering \includegraphics[width=1.0\linewidth]{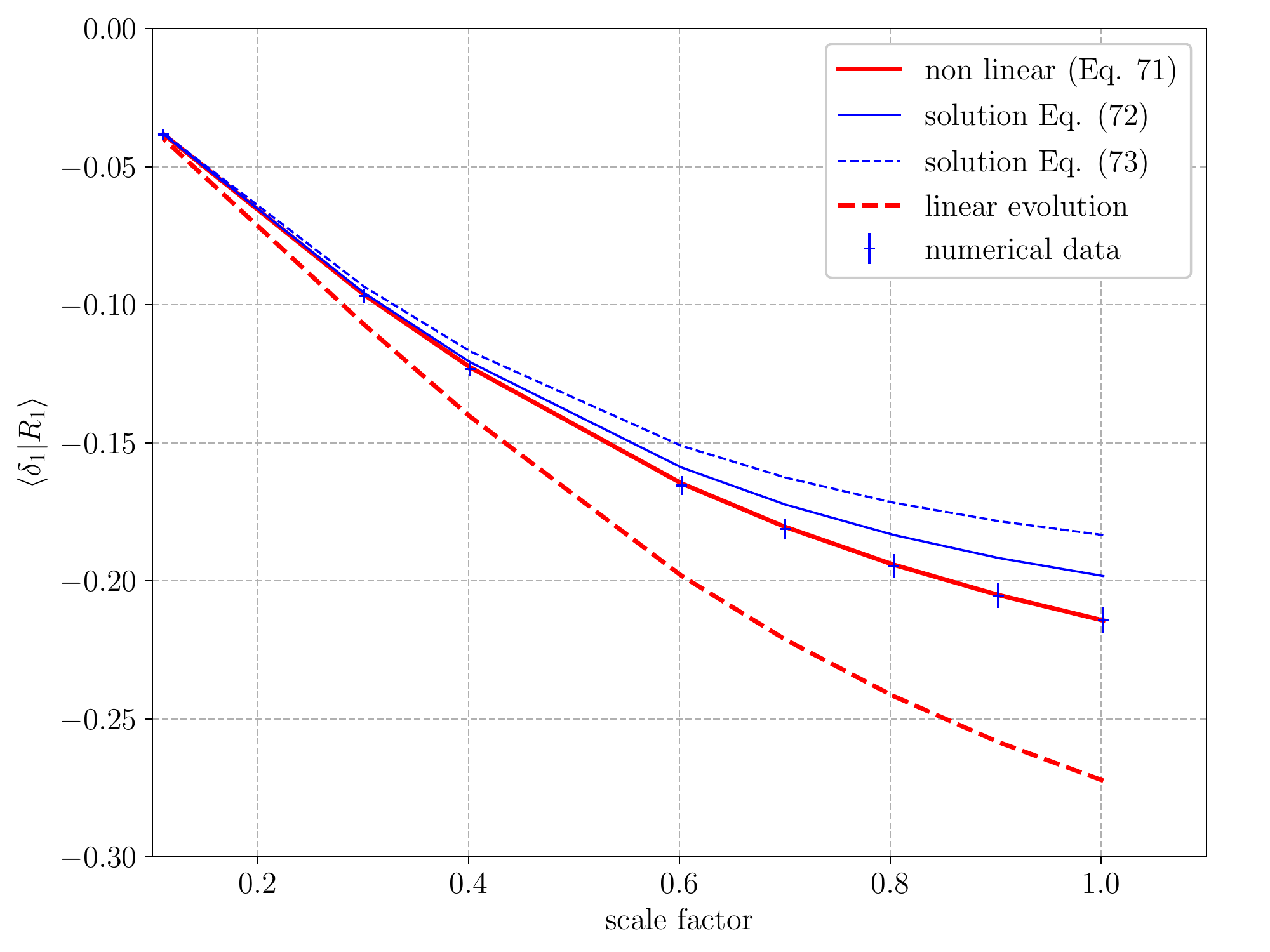}
    \caption{Redshift evolution of the first conditional moment of $\delta_1$ at fixed $R_1$,\ie $\avg{\delta_1|R_1}$ for $R_1=20$ \Mpc. The red curve is the non linear solution derived from Lagrangian spherical dynamics \refeq{mean_d1_of_t}, the dashed red curve is the linear solution $\avg{\delta_1|R_1}(t)=D(t)/D(t_i)\avg{\delta_1|R_1}$ and the numerical data are in blue (points with corresponding error bars). The agreement between numerical and theoretical computation from \refeq{mean_d1_of_t} is very good for all redshift (here the $x$ axis is the scale factor $a=1/(z+1)$. The full blue line and the dashed line are the small $\delta_1$ expansion from \refeq{d1_expansion} and \refeq{d1_expansion_wrong}. At low redshift, all the non linear terms beyond the second order term $\avg{\delta_1|R_1}$ have to be taken into account to reproduce the numerical results.}
        \label{fig:d1_evolution}
\end{figure}

In \refapp{app:d1_moments} we show that for both central minima and central maxima, these moments can be simply rewritten in term of  the primordial moments in Gaussian random field

\begin{equation}
\avg{\delta_1^n|R_1}(t)=\sum_{k\geq 0} \tilde{D}(t)^n\left(1-\tilde{D}(t)\right)^k\binom{-n}{k}\avg{\delta_1^{n+k}|R_1}
\end{equation}
In particular, for $n=1$ we get
\begin{equation}
\label{mean_d1_of_t}
\avg{\delta_1|R_1}(t)=\tilde{D}(t)\sum_{k\geq 0}\left(\tilde{D}(t)-1\right)^k\avg{\delta_1^{1+k}|R_1}
\end{equation}
At $t=t_i$, since $\tilde{D}(t_i)=1$, the only non zero contribution comes from the $k=1$ term leading to $\avg{\delta_1|R_1}$. Expanding \refeq{mean_d1_of_t} we have
\begin{equation}
\label{d1_expansion}
\avg{\delta_1|R_1}(t)\simeq \tilde{D}\avg{\delta_1|R_1} + \tilde{D}(\tilde{D}-1)\avg{\delta_1^2|R_1} + ...
\end{equation}
The first term is the linear evolution while higher terms account for the corrections to this simple dynamics. Note that \refeq{mean_d1_of_t} is different from the evolution of the mean which would be
\begin{equation}
\label{dtfalse}
\avg{\delta_1|R_1}(t)= \frac{ \avg{\delta_1|R_1}\tilde{D}(t)}{1 - \avg{\delta_1|R_1}\left(\tilde{D}(t)-1\right)} 
\end{equation}
whose small $\tilde{D}(t)$ expansion is
\begin{equation}
\label{d1_expansion_wrong}
\avg{\delta_1|R_1}(t)\simeq \tilde{D}\avg{\delta_1|R_1} + \tilde{D}(\tilde{D}-1)\avg{\delta_1|R_1}^2 + ...
\end{equation}
The first linear term remains unchanged while the second one differs by $\avg{\delta_1^2|R_1}-\avg{\delta_1|R_1}^2$. In \refim{fig:d1_evolution} we show the measure of $\avg{\delta_1|R_1}(t)$ in the numerical simulations together with the exact evolution \refeq{mean_d1_of_t} and the various approximations \refeq{d1_expansion} and \refeq{d1_expansion_wrong} for $R_1=20$ \Mpc. We also show the linear prediction $\avg{\delta_1|R_1}=\tilde{D}(t)\avg{\delta_1|R_1}$. It turns out that the non linear prediction fits very well the measured evolution on the whole range of reshifts.

The possibility to predict precisely the distribution of the compensation density at any non linear redshift opens again new possibilities for cosmology and will be deeply studied in \citet{paper3,paper4}

%%%%%%%%%%%%%%%%%
%%%%   SECTION  4  %%%%%
%%%%%%%%%%%%%%%%%
\section{Discussion and conclusion}

In this paper, we derived the main statistical properties of CoSpheres as introduced in \paperI both in the primordial GRF and in the structured Universe until $z=0$. 

Within the Gaussian field, CoSpheres are fully determined by a unique compensation radius and a set of shape parameters $\nu$, $x$ and $\nu_1$. This formalism can be seen as a physical extension of the original BBKS work by taking explicitly into account the large scale matter field around the local extremum. This extension describes the correlation between local extremum and their large scale environment.

In the framework of GRF, we derive the full joint Gaussian probability for the four parameters $R_1$, $\nu$, $x$ and $\nu_1$ \refeql{Pfull} by taking into account the appropriate domain for the curvature parameter $x$ in order to insure the correct definition of $R_1$ \refsecl{sec:FCC}. Interestingly, as studied in \refsec{sec:R1_infty}, the very large scale limit $R_1\to +\infty$ reduces to the standard BBKS statistics for the central extrema \citep{BBKS}. Physically, it describes the limit where the central extrema is completely decorrelated from its surrounding cosmic environment. In other words, The statistical distribution of $\nu$ or $x$ are no more affected by $R_1$ when $R_1$ becomes very large.

\begin{figure}
	\includegraphics[width=1.0\linewidth]{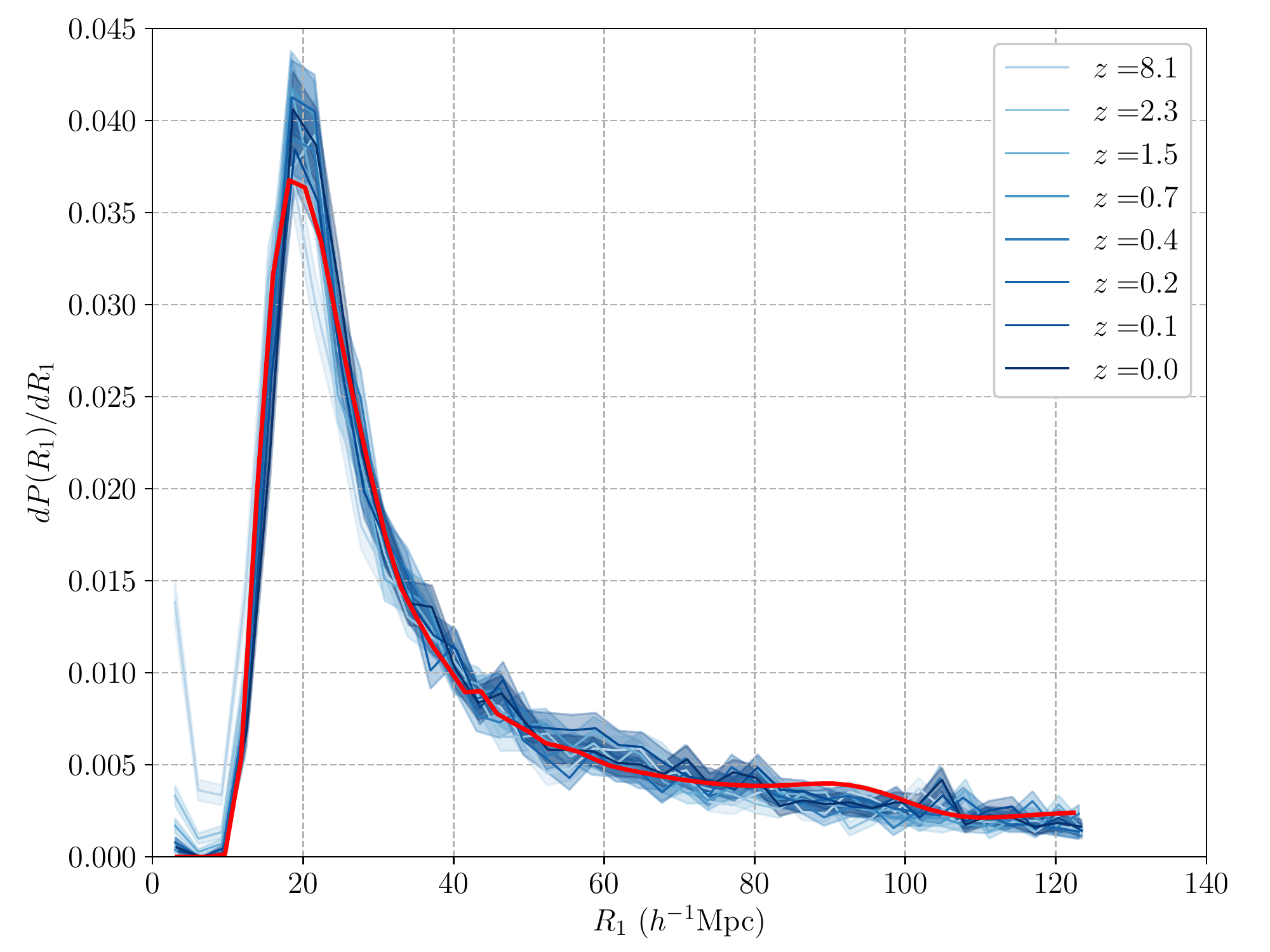}
	\caption{Probability distribution $dP(R_1)/dR_1$ of the compensation radius $R_1$ from $z=8$ to $z=0$ centered on local minima. For this plot, the density field has been smoothed on a Gaussian scale $R_f=5$ \Mpc. This figure has been obtained from $10000$ voids without selection criteria except that the central density contrast is negative. The shaded region around each curve is the Poisson noise computed in radial bins of size $dR_1=3.5$ \Mpc. The red curve is the analytical prediction computed in the primordial Gaussian conditions \refeq{R1_distribution} where the power-spectrum has been smoothed on the equivalent Gaussian radius $R_g=5$ \Mpc}
	\label{fig:PR1_z0_theory_void}
\end{figure}

Marginalizing the full joint probability over the shape parameters $\nu$, $x$ and $\nu_1$ leads to the distribution $dP(R_1)$ \refeql{R1_distribution} which gives the probability to find a CoSphere with a given $R_1$. Since each single $R_1$ is a comoving quantity, its pdf $dP(R_1)$ is also expected to be conserved in comoving coordinates during the whole cosmic evolution. This is confirmed by \refim{fig:PR1_z0_theory} where we compare the $R_1$ distribution around DM haloes (central extremum) at various redshifts with the Gaussian prediction (red curve). Since the Gaussian field is exactly symmetric, this distribution can also be transposed without any change to the complementary case of central minimum, seeding cosmic voids. In \refim{fig:PR1_z0_theory_void} we show the compensation radius distribution $dP(R_1)$ at various redshift for central minima. This figure has been obtained by finding minimum in the density field smoothed with a Gaussian kernel on $R_f=5$ \Mpc at $z=0$ and assuming that their position do not change with redshift (profiles are computed around the same position for each $z$). Once again, the Gaussian prediction (red curve) fits the measured distribution on all available scales.

As in \refim{fig:PR1_z0_theory}, the BAO-like wiggles around $R_1\sim 90$ \Mpc expected from theory do not appears clearly on numerical data. As previously discussed in \refsec{SizeEvolution}, this slight discrepancy between theoretical and numerical results may be due to the cosmic variance which dominates on this scales due to the size of our simulation box \citep{Rasera2013}.

We emphasize that this distribution is suited to model the distribution of cosmic void sizes once identified as spherically compensated regions. This approach is fundamentally different from other attempts to model void statistics such as in \citet{Sheth2004a, Furlanetto2006, Achitouv2015}. These approach are based on the excursion set theory \cite{Press1974, Bond1991} while our formalism identifies the size of a void to its compensation radius. The improvement of our approach is the ability to define correctly the size of such cosmic structure and to be able to model its properties from first principles. However, our model assumes that we are indeed able to find this radius in observable data, which is far from being obvious.

Apart from the compensation radius distribution, we derived the statistical properties of the shape parameters of CoSpheres and particularly the conditional probability distribution of each shape parameter at fixed $R_1$. We computed their conditional moment and discussed the correlation between the central peak and its surrounding cosmic environment. More precisely, we have shown that whilst $R_1$ increases, the peak parameters $\nu$ and $x$ progressively tend to their asymptotic BBKS value while $\nu_1$ vanishes. Small central extremum (small value of $\abs{\nu}$) are associated with narrow compensation radii with a high compensation density $\nu_1$. On the other hand, higher peaks are more likely to sit in large inhomogeneous regions with a small compensation density. Once again, this discussion is valid for both central maximum and minimum, describing cosmic voids.
 
Using the spherical collapse model and the conservation of $R_1$ we then derived the evolved conditional distribution for $\delta_1$ at fixed $R_1$. This leads to the evolved moments $\avg{\delta_1^n|R_1}$ at $z=0$ whose expression can be computed analytically. The comparison with numerical simulation are in a very good agreement with the Lagrangian prediction (see \refim{fig:PR1_z0_theory} and \refim{fig:distribution_d1_today}). In the opposite the Eulerian dynamical evolution fails to reproduce these quantities, even on "large scales", e.g. $R_1=40$ \Mpc. 

The statistical properties of the compensation scalars $R_1$ and $\delta_1$ are thus particularly interesting since they can be directly measured in numerical simulations or otherwise from observational data and can be used as new cosmology probes. This is investigated in \citet{paper3} and \cite{paper4}. 

The fundamental interest of CoSpheres for cosmology is thus based on two main properties. The first one is the conservation of the compensation radius $R_1$ in comoving coordinates, \ie the fact that that $R_1(t)\propto a(t)$. This fundamental feature implies the conservation of its probability distribution $dP(R_1)$ during the whole cosmic history and allows in principle to probe directly the properties of the primordial Gaussian Universe. This property allows to evaluate at $z=0$ the statistics of the shape parameters describing both the small scale extremum and its large scale environment. The second fundamental property is the exact symmetric treatment of CoSpheres defined from central maximum or minimum. This formalism provides a physically motivated model for cosmic voids and offers an alternative approach for describing their statistical properties.

\bibliographystyle{mnras}
\bibliography{biblio}

%%%%%%%%%%%%%%%%
%%%%  APPENDIX  %%%%%
%%%%%%%%%%%%%%%%
\appendix

%%%%%%%%%%%%%%%%%%%%%%%%
\section{Coefficients $C_\alpha$}\label{Calpha}
In this appendix we give the explicit expressions of the $C_\alpha$ coefficients appearing in \refeq{F}
\begin{align}
\nonumber \frac{C_x}{\avg{k^4}}=-\J^2\WW &+ 2\J\W\WJ - \WJ^2\\
&+ \JJ\left[\WW - \W^2\right]
\end{align}

\begin{align}
\nonumber C_{\nu} = &-\JJ\kkW^2  - \kkkk\WJ^2\\
\nonumber & + 2\kkJ\kkW\WJ \\
& + \WW\left[\JJ\kkkk - \kkJ^2\right]
\end{align}

\begin{align}
\nonumber C_{\nu_1} = -\kkW^2 &+ 2\kk\kkW\W - \kk^2\WW\\
&+ \kkkk\left[\WW - \W^2\right]
\end{align}

\begin{align}
\nonumber \frac{C_{x\nu}}{\sqrt{\kkkk}} =  & \JJ\kkW\W - \JJ\kk\WW + \J\kkJ\WW\\
&+ \WJ \left[\kk\WJ - \J \kkW - \W\kkJ \right]
\end{align}

\begin{align}
\nonumber \frac{C_{x\nu_1}}{\sqrt{\kkkk}} = &\kkJ\W^2 - \J\kkW\W\\ 
\nonumber &+\WW\left[\J\kk-\kkJ\right]\\
&+ \WJ\left[\kkW - \W\kk\right]
\end{align}

\begin{align}
\nonumber C_{\nu_1\nu} = & \kk\kkJ\WW - \kkJ\kkW\W \\
\nonumber & + \J\left[\kkW^2- \kkkk\WW\right] \\
 & + \WJ\left[\kkkk\W - \kk\kkW\right]
\end{align}
We also introduce $C_0$ defined by
\begin{align}
\label{C0}
\nonumber C_0 = & \left(\J \kkW - \kkJ \W\right)^2\\
& + \left(\kk^2 - \kkkk\right)\times\left(\JJ\WW - \WJ^2\right)
\end{align}
Linked to the determinant of the correlation sub-matrix $\tilde{\boldsymbol{Q}}$
\begin{equation}
\Sigma^2(R_1) = C_0 + C_x + C_\nu + 2\frac{\kk}{\sqrt{\kkkk}}C_{x\nu}
\end{equation}
Note that all these coefficients are functions of $R_1$.

%%%%%%%%%%%%%%%%%%

%%%%%%%%%%%%%%%%%%%
\section{Computing the evolved moments of the compensation density}
\label{app:d1_moments}
We now compute the evolved moments $\avg{\delta_1^n|R_1}(t)$ for both central minimum and maximum. We show that in both cases it gives
\begin{equation}
\label{moments_today}
\avg{\delta_1^n|R_1}(t)=\sum_{k\geq 0} \tilde{D}(t)^n\left(1-\tilde{D}(t)\right)^k\binom{-n}{k}\avg{\delta_1^{n+k}|R_1}
\end{equation}
where the various moments $\avg{\delta_1^m|R_1}$ are computed within the Gaussian field at some time $t_i$ and $\tilde{D}(t)=D(t)/D(t_i)$.

%--------------------------------
\subsection{Central minima, \ie cosmic voids}
For central minimum seeding cosmic voids, the compensation density $\delta_1$ is positive. Using the notations of \refsec{sec:d1_dynamics}, $\delta_1^t$ is the evolved compensation density and $\delta_1$ its corresponding value in the primordial field. These quantities are linked through \refeq{deltat} and \refeq{delta0}. Since we consider finite values of $\delta_1^t$ today, this implies that the corresponding primordial values must satisfy $\delta_1\leq \delta_1^c(t)\equiv 1/(\tilde{D}(t) - 1)$ \refeql{deltat}. The moments today are given by 
\begin{equation}
\avg{\delta_1^n|R_1}(t)=\int_0^{+\infty}\left(\delta_1^t\right)^ndP(\delta_1^t|R_1)
\end{equation}
Using the mapping \refeq{deltat} it leads to
\begin{align}
\avg{\delta_1^n|R_1}(t)&=\int_0^{\delta_1^c(t)}\left(\frac{\tilde{D}(t)\delta_1}{1- \delta_1(\tilde{D}(t)-1)}\right)^ndP(\delta_1|R_1)\\
&=\left(\sigma_0\tilde{D}(t)\right)^n\int_0^{1/\epsilon(t)}\left(\frac{\nu_1}{1- \nu_1\epsilon(t)}\right)^ndP(\nu_1|R_1)
\end{align}
where $\delta_1=\sigma_0 \nu_1$ and $\epsilon(t)=\sigma_0(\tilde{D}(t) - 1)$. Since $\nu\in[0, 1/\epsilon(t)]$, we can use a Maclaurin expansion of the $1/(1-\nu_1\epsilon(t))$ term, leading to 
\begin{align}
\nonumber\avg{\delta_1^n|R_1}(t)=\tilde{D}(t)^n&\sum_{k\geq 0}(-1)^k\left(\tilde{D}(t)-1\right)^k \binom{-n}{k}\\
&\times\int_0^{1/\epsilon(t)}\left(\sigma_0\nu_1\right)^{n+k}dP(\nu_1|R_1)
\end{align}
Using the primordial moments $\avg{\delta_1^n|R_1}=\sigma_0^n\avg{\nu_1^n|R_1}$ we get the final result as recalled in \refeq{moments_today}.

%--------------------------
\subsection{Central maximum}
For central maximum, the computation necessitates a careful treatment. We start from 
\begin{equation}
\label{moments_0}
\avg{\delta_1^n|R_1}(t)=\int_{-1}^0\left(\frac{\tilde{D}(t)\delta_1}{1- \delta_1(\tilde{D}(t)-1)}\right)^ndP(\delta_1|R_1)
\end{equation}
Using \refeq{delta0}, it is clear that if $\delta_1\geq -1$, then for any time $t$ we have $\delta_1^t\geq -1$. However, we cannot use here a Maclaurin expansion since the term $\nu_1\epsilon(t)= \delta_1(\tilde{D}(t)-1))$ is no more included in its convergence radius, \ie it can take values larger than $1$ ($\abs{\nu_1\epsilon(t)}>1$). We thus introduce the new variables $x=\delta_1+1$ and $\eta(t)=(\tilde{D}(t) - 1)/\tilde{D}(t)$, both included in $[0, 1]$. \refeq{moments_0} transforms to
\begin{equation}
\avg{\delta_1^n|R_1}(t)=\int_0^1\left(\frac{x-1}{1-x\eta(t)}\right)^ndP(x-1|R_1)
\end{equation}
after a Taylor expansion in term of $\eta(t)$ and switching back to $\delta_1$ we get
\begin{equation}
\avg{\delta_1^n|R_1}(t)=\sum_{k\geq 0}\left(-\eta(t)\right)^k\binom{-n}{k}\avg{\delta_1^n(1+\delta_1)^k|R_1}
\end{equation}
Since $\delta_1\in[-1, 0]$, we expand also the term $(1+\delta_1)^k$ term, 
\begin{equation}
\avg{\delta_1^n|R_1}(t)=\sum_{k\geq 0}\left(-\eta(t)\right)^k\binom{-n}{k}\sum_{p=0}^k\binom{k}{p}\avg{\delta_1^{n+p}|R_1}
\end{equation}
We simplify this expression by reordering and collecting terms with the same contribution, 
\begin{equation}
\avg{\delta_1^n|R_1}(t)=\sum_{m\geq 0}\left[\sum_{k\geq m} (-\eta(t))^k\binom{-n}{k}\binom{k}{m}\right]\avg{\delta_1^{m+n}|R_1}
\end{equation}
Using again $\eta(t)=(\tilde{D}(t)-1)/\tilde{D}(t)$ together with the relation
\begin{equation}
\forall \alpha\in[-1, 1], \quad\sum_{k\geq m} \alpha^k\binom{-n}{k}\binom{k}{m}=\frac{\alpha^m}{(1+\alpha)^{m+n}}\binom{-n}{m}
\end{equation}
where $\alpha=(1-1/\tilde{D}(t))=-\eta(t)$. We finally recover the same expression \refeq{moments_today} which holds for both central minima and central maxima.

\end{document}